\documentclass[11pt,a4paper]{article}

\usepackage{epsfig}
\usepackage{amsmath}
\usepackage{amssymb}
\usepackage{verbatim}
\usepackage{bm}
\usepackage{dsfont}
\usepackage[footnotesize]{caption}
\usepackage{subfigure}
\usepackage{cite}
\usepackage{textcomp}
\usepackage{calc}
\usepackage{geometry}
\usepackage{bbm}
\usepackage{xcolor}
\usepackage[utf8]{inputenc}
\def\be{\begin{equation}}
\def\ee{\end{equation}}
\def\ba{\begin{eqnarray}}
\def\ea{\end{eqnarray}}
\usepackage{ulem}

\usepackage{hyperref}
\hypersetup{colorlinks=true,urlcolor=blue,linkcolor=red,citecolor=green!60!black}

\renewcommand{\baselinestretch}{1.3}

\begin{document}


\noindent February 2020
\hfill DESY 20-017

\vskip 1.5cm

\begin{center}

\bigskip
{\huge\bf Resonant backreaction 
in axion inflation}

\vskip 2cm

\renewcommand*{\thefootnote}{\fnsymbol{footnote}}

{\large
Valerie Domcke$^{a}$,%
Veronica Guidetti$^{b,\,c}$,%
Yvette Welling$^{a}$\,%
and Alexander Westphal$^{a}$
}\\[3mm]
{\it{
$^{a}$ Deutsches Elektronen-Synchrotron (DESY), 22607 Hamburg, Germany\\
$^{b}$  Dipartimento di Fisica e Astronomia, Universit\`a di Bologna, \\ via Irnerio 46, 40126 Bologna, Italy\\
$^{c}$ INFN, Sezione di Bologna, viale Berti Pichat 6/2, 40127 Bologna, Italy}}

\end{center}

\vskip 1cm

\renewcommand*{\thefootnote}{\arabic{footnote}}
\setcounter{footnote}{0}


\begin{abstract}
Axion inflation entails a coupling of the inflaton field to gauge fields through the Chern-Simons term. This results in a strong gauge field production during inflation, which backreacts on the inflaton equation of motion. Here we show that this strongly non-linear system generically experiences a resonant enhancement of the gauge field production, resulting in oscillatory features in the inflaton velocity as well as in the gauge field spectrum. The gauge fields source a strongly enhanced scalar power spectrum at small scales, exceeding previous estimates. For appropriate parameter choices, the collapse of these over-dense regions can lead to a large population of (light) primordial black fholes with remarkable phenomenological consequences.
\end{abstract}

\thispagestyle{empty}


\newpage

{\hypersetup{linkcolor=black}\renewcommand{\baselinestretch}{1}\tableofcontents}

\newpage

\section{Introduction}

Axion-like particles are among the prime candidates for particle physics implementations of cosmic inflation. Protected by an approximate shift-symmetry, these Pseudo Nambu Goldstone Bosons naturally come with a sufficiently flat scalar potential to support slow-roll inflation. Many concrete realizations of axion inflation in field theory have been proposed beginning with Ref.~\cite{Freese:1990rb}, for axions in string theory see~\cite{Banks:2003sx,Svrcek:2006yi}. 

The shift-symmetry of the axion-like inflaton $\Phi$ allows for a derivative coupling to the field strength tensor $F_{\mu \nu}$ of a (dark) gauge sector,
\begin{equation}
 {\cal L}_\text{int} = - \frac{\sqrt{-g}}{4 f} \Phi F_{\mu \nu} \tilde F^{\mu \nu} \,,
 \label{eq:int}
\end{equation}
with $f$ denoting the axion decay constant and for simplicity, we will consider $F_{\mu \nu}$ to describe a hidden sector abelian gauge group, i.e.\ a dark photon.\footnote{If the theory contains particles charged under this $U(1)$ (as is e.g.\ the case for the Standard Model hypercharge), these particles must be included in the analysis if they are sufficiently light, as they will be produced via Schwinger production from the vacuum, thereby significantly damping the gauge field production. On the contrary, the impact of heavier particles is exponentially suppressed and they can be safely integrated out~\cite{Domcke:2018eki,Domcke:2019qmm}.} This interaction triggers a tachyonic instability of the dark photon driven by the velocity $\dot \Phi$ of the inflaton, leading to an exponential production of dark photons~\cite{Turner:1987vd,Garretson:1992vt,Anber:2006xt}. The resulting non-thermal gauge field distribution backreacts on the inflaton, dampening its motion. At the same time, the gauge fields act as a source of scalar and tensor perturbations~\cite{Barnaby:2010vf,Barnaby:2011qe,Barnaby:2011vw,Cook:2011hg}, in addition to the standard vacuum fluctuations amplified during cosmic inflation. These perturbations can be probed by CMB observations~\cite{Barnaby:2010vf,Meerburg:2012id}, searches for primordial black holes~\cite{Linde:2012bt, Garcia-Bellido:2016dkw,Domcke:2017fix,Cheng:2018yyr} and gravitational wave experiments~\cite{Cook:2011hg,Anber:2012du,Domcke:2016bkh,Bartolo:2016ami}, rendering axion inflation not only a theoretically well motivated but also an experimentally testable proposal for cosmic inflation~\cite{Barnaby:2011qe}.

In this work we have a closer look at the backreaction of the gauge field distribution on the inflaton equations of motion. Since this determines the evolution of the homogeneous inflaton field, this has a crucial impact on all potential observables of this framework. The interaction~\eqref{eq:int} results in a friction term in the background equation of motion for $\Phi$ which is proportional to $\langle F \tilde F \rangle$. In Fourier space, this non-linear interaction involves an integral over all relevant Fourier modes of the gauge field, leading to a integro-differential system describing the evolution of the gauge field modes and the homogeneous component of the inflaton.

In many previous works, this system is solved by assuming the inflaton velocity to be constant in the gauge field equation of motion (see e.g.~\cite{Barnaby:2011qe}), motivated by the usual slow-roll approximation employed in cosmic inflation. However, since the gauge field enhancement and hence the backreaction on the inflaton are exponentially sensitive to this velocity, this approximation becomes invalid in the phenomenologically interesting regime of sizable gauge field production. Recently, several alternative approaches have been put forward. Lattice simulations~\cite{Adshead:2015pva,Cuissa:2018oiw,Adshead:2019lbr}, focusing mainly on the preheating phase, accurately capture the backreaction but are limited in the amount of time evolution that can be tracked. Ref.~\cite{Sobol:2019xls} proposed a gradient expansion of the generated electric and magnetic field.
Self-consistent numerical solutions of the integro-differential system have been obtained in Refs.~\cite{Cheng:2015oqa,Notari:2016npn,DallAgata:2019yrr}. These latter studies noted the appearance of remarkable oscillatory features in the inflaton velocity. In this work, we reproduce these findings and quantitatively explain the occurring resonance phenomenon based on semi-analytical arguments.
Since the enhancement of the gauge field modes is most sensitive to the inflaton velocity around horizon crossing whereas the backreaction is dominated by super-horizon gauge field modes, the system responds with a time delay to a change in the inflaton velocity. This time delay is logarithmically sensitive to the inflaton velocity. As the inflaton velocity increases during the course of inflation the system hits its resonance frequency, leading to strong oscillations in the amplitude of $\langle F \tilde F \rangle$ as a function of time. This crucially impacts both the background equation of motion as well as the generation of scalar and tensor perturbations.

The power spectrum of scalar perturbations can be obtained by solving the linearized inhomogeneous equation of motion for the inflaton field taking into account the backreaction and source terms proportional to $F \tilde F$. In the pioneering works~\cite{Anber:2009ua,Barnaby:2010vf,Barnaby:2011vw,Linde:2012bt} this task has been solved in the weak and very strong backreaction regime. Here we extend these results to arbitrary inflaton gauge field couplings by numerically determining the Greens function including the backreaction term. We report two important results. Firstly, for a smoothly growing $\langle F \tilde F \rangle$, we find that the analytical estimate in~\cite{Anber:2009ua} significantly overestimates the backreaction compared to our full numerical results. As a result, the actual power spectrum is significantly enhanced compared to previous estimates. Consequently, a large primordial black hole (PBH) abundance can be generated, leading to an early PBH dominated phase. Requiring the transition to radiation domination to occur before the onset of big bang nucleosynthesis imposes stringent constraints on the parameter space. Secondly, for an oscillating $\langle F \tilde F \rangle$ as found in the numerical solution of the background equation of motion, the scalar power spectrum features prominent peaks which, for suitable parameters, may lead to a PBH population peaked at logarithmically equidistant masses, accompanied by a gravitational wave spectrum with similar features. This would be a smoking gun signature of the resonance phenomenon inherent to axion inflation.

The remainder of this paper is organized as follows. In Sec.~\ref{sec:axioninflation} we review the mechanism of axion inflation. Sec.~\ref{sec:toymodel} explains the resonance inherent to this coupled system of differential equations and provides analytical estimates for the relevant time scales, {which are further refined in appendix \ref{app:phaseshift}.} This is numerically confirmed by our numerical results presented in Sec.~\ref{sec:numerics} for two exemplary values of the axion decay constant. Based on these results for the background evolution, we compute the power spectrum of scalar fluctuations in Sec.~\ref{sec:PS} before concluding in Sec.~\ref{sec:conclusions}. Details on our numerical procedure as well as on the comparison with previous works can be found in appendices~\ref{app:numerics} and \ref{app:comparison}, respectively.

\section{Axion inflation \label{sec:axioninflation}}

We consider a pseudo-scalar $\Phi$ coupled to the field strength tensor $F_{\mu \nu}$ of an abelian gauge group through a shift-symmetric coupling {(see e.g.~\cite{Barnaby:2011qe} for a review)},
\begin{align}
 \frac{{\cal L}}{\sqrt{-g}} = - \frac{1}{2} \partial_\mu \Phi \partial^\mu \Phi - \frac{1}{4} F_{\mu \nu} F^{\mu \nu} - V_{,\Phi} - \frac{1}{4 \, f } \Phi F_{\mu \nu} \tilde F^{\mu \nu} \,.
\end{align}
Here $V(\Phi)$ is a scalar potential explicitly breaking the shift-symmetry of $\Phi$ and $\tilde F^{\mu \nu} = \epsilon^{\mu \nu \rho \sigma} F_{\rho \sigma}/(2 \sqrt{-g})$ with $\epsilon^{0123} = 1$ is the dual field strength tensor. Working in quasi de-Sitter space we introduce the time variable
\begin{equation}
 N = \int H dt \,,
\end{equation}
where $H = \dot a/a$ denotes the (approximately constant) Hubble parameter. In the separate Universe picture, the number of e-folds $N$ elapsed in a time interval $[t_1, t_2]$ between two equal-density hyper surfaces varies by $\delta N$ between `separate', locally homogeneous universes, accounting for the inhomogeneities in our primordial Universe~\cite{Starobinsky:1986fxa,Salopek:1990jq,Sasaki:1995aw,Yokoyama:2007uu}. Expanding\footnote{{Here we are dropping terms of ${\cal O}(\delta N^2)$, assuming $\delta N \ll 1$. Moreover, throughout this  paper, we will neglect the spatial gradients of the inflaton field. As we will see later, due to the strong enhancement of the scalar power spectrum in axion inflation, this is a non-trivial limitation of our analysis. To go beyond this and include strong spatial gradients of the scalar and gauge field into the analysis would require moving beyond the $\delta N$-formalism, e.g. along the lines of the full quantum formalism of~\cite{Gorbenko:2019rza}.}} 
\begin{align}
 \Phi = \Phi_{\delta N = 0} + \frac{\partial \Phi}{\partial N} \bigg|_{\delta N = 0} \delta N \equiv \phi + \delta \phi
\end{align}
we obtain the equation of motion for the homogeneous part
\begin{align}
 \phi'' + \frac{H'}{H} \phi' + 3 \phi' + \frac{V_{,\phi}}{H^2} - \frac{1}{f H^2} \, \langle \vec E \vec B \rangle = 0 \,,
 \label{eq:eom_phi_0}
\end{align}
with $' = \partial/\partial N$ and $\langle \dots \rangle$ denoting the average over many universes, thus selecting the globally homogeneous contribution.\footnote{Here we assume a definite sign for {the initial value of} $\phi'$. {In a $CP$ conserving universe this corresponds to averaging over a finite subset of Hubble patches.}}

Turning to the gauge fields, the $CP$-odd nature of $F_{\mu \nu} \tilde F^{\mu \nu}$ will be most transparent when expanding in Fourier-modes of the comoving vector potential in the chiral basis,
\begin{align}
 \vec A(\tau, \vec x) = \int \frac{d^2 k}{(2 \pi)^{3/2}} \sum_{\sigma = \pm} \left[ A_\sigma (\tau, \vec k) \hat e_\sigma(\hat k) \hat{\bm a}(\vec k) e^{i \vec k \vec x} + A_\sigma^* (\tau, \vec k) \hat e^*_\sigma(\hat k) \hat{\bm a}^\dagger(\vec k) e^{-i \vec k \vec x} \right]\,,
\end{align}
with the polarization tensors obeying $\hat e_\sigma(\hat k) \cdot \vec k = 0$, $\hat e_\sigma(\hat k) \cdot \hat e_{\sigma'}(\hat k) = \delta_{\sigma \sigma'}$ and $i \vec k \times \hat e_\sigma(\hat k) = \sigma k \hat e_\sigma(\hat k)$ where $\vec k = |\vec k| \hat k = k \, \hat k$, $\hat{\bm a}$ ($\hat{\bm a}^\dagger$) denoting the annihilation (creation) operators and $d \tau = dt/a$ denoting conformal time. In this basis, the equation of motion for the Fourier coefficients $A_\sigma(\tau, \vec k)$ is obtained as
\begin{align}
 \frac{d^2 A_\pm(\tau, \vec k)}{d \tau^2} + \left[k^2 \pm 2 \lambda \xi k a H \right]A_\pm(\tau, \vec k) = 0  \qquad \text{with} \quad \xi \equiv \frac{\lambda \phi'}{2 f} > 0\,, \label{eq:eom_A}
\end{align}
where $\lambda \equiv \text{sign}(\phi')$. 
For a sufficiently large inflaton velocity the effective mass term in the square brackets for the helicity mode with $\sigma = - \lambda$ undergoes a tachyonic instability, leading to an exponential enhancement. These gauge fields backreact on the inflaton equation of motion. The physical electric and magnetic fields entering in~\eqref{eq:eom_phi_0} are obtained as
\begin{align}
 \vec E = - \frac{1}{a^2} \frac{d \vec A}{d \tau} \,, \qquad \vec B = \frac{1}{a^2} \vec \nabla \times \vec A \,,
\end{align}
leading to
\begin{align}
 \langle \vec E \vec B \rangle = - \frac{\lambda}{a^4} \int \frac{d k}{4 \pi} k^3 \frac{d}{d \tau}  \left| A_{- \lambda}(\tau, \vec k)  \right|^2 \,, \label{eq:EB}
\end{align}
and the energy density
\begin{align}
 \left\langle \frac{E^2+B^2}{2} \right\rangle = \frac{1}{a^4} \int \frac{d k}{4 \pi^2} \, k^2\left(   \left|\frac{d  A_{- \lambda}(\tau, \vec k) }{d \tau} \right|^2 +k^2   \left|  A_{- \lambda}(\tau, \vec k) \right|^2\right)\,, \label{eq:EEBB}
\end{align}
where we have considered only the dominant, enhanced helicity mode. In summary, Eqs.~\eqref{eq:eom_phi_0}, \eqref{eq:eom_A} and \eqref{eq:EB}, together with the Friedmann equation
\begin{align}
 3 H^2 M_P^2 = V(\phi) + \frac{1}{2} H^2 (\phi')^2 + \left\langle \frac{E^2+B^2}{2} \right\rangle  \,,
\end{align}
form a closed, integro-differential system of equations describing the gauge field production induced by the motion of the inflaton, taking into account the backreaction of these gauge fields.

\section{Resonant gauge field production \label{sec:toymodel}}

In the limit of quasi de-Sitter space-time, $\tau = -1/(a H)$, and for constant $\xi$, Eq.~\eqref{eq:eom_A} can be solved exactly. For the enhanced mode, this yields
\begin{align}
 A_{-\lambda}(\tau, \vec k) = \frac{e^{\pi \xi/2}}{\sqrt{2k}} W_{-  i \xi, 1/2}(2 i k \tau) \,.
\label{eq::AkWhit}
\end{align}
Here $W_{k,m}(z)$ denotes the Whittaker function and we have imposed Bunch Davies vacuum as an initial condition for far sub-horizon modes. Inserting this into Eqs.~\eqref{eq:EB} and \eqref{eq:EEBB} yields
\begin{align}
 \langle \vec E \vec B \rangle \simeq -  \frac{\lambda e^{2 \pi \xi}}{2^{21} \pi^2 \xi^4} H^4 \int_0^{x_\text{uv}} x^7 e^{-x} dx \simeq - 2.4 \cdot 10^{-4} \, \lambda \, H^4 \frac{e^{2 \pi \xi}}{\xi^4} \,,
 \label{eq:EB_estimate}
\end{align}
and
\begin{align}
 \left\langle \frac{E^2+B^2}{2} \right\rangle \simeq  \frac{ e^{2 \pi \xi}}{2^{19} \pi^2 \xi^3} H^4 \left[\int_0^{x_\text{uv}} x^6 e^{-x} dx +\frac{1}{(2^3\xi)^2}\int_0^{x_\text{uv}} x^8 e^{-x} dx\right] \simeq 1.3 \cdot 10^{-4} \, H^4 \frac{e^{2 \pi \xi}}{\xi^3
} \,,
 \label{eq:EEBB_estimate}
\end{align}
with $x_\text{uv} \simeq 2 \xi$ ensuring the cut-off of the UV divergence. The last equality is valid for $\xi \gtrsim 3$, smaller values of $\xi$ require a more careful regularization scheme~\cite{Jimenez:2017cdr,Ballardini:2019rqh}.

\begin{figure}[t!]
    \centering
    \includegraphics[width=0.6\textwidth]{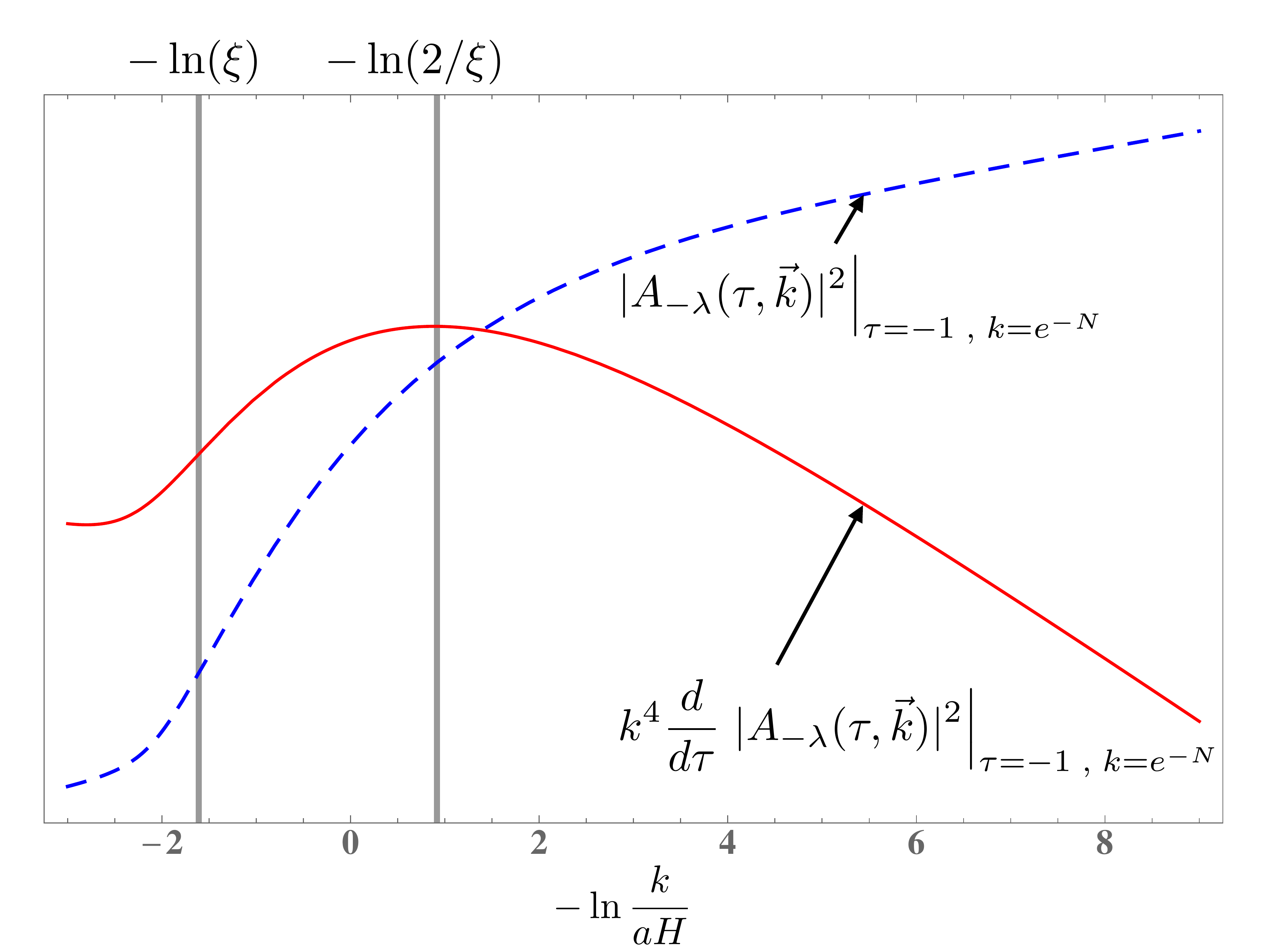}
    \caption{Blue dash: The square of the gauge field mode  $|A_{-\lambda}(\tau,\vec k)|^2$. Red solid: The $ \langle \vec E \vec B \rangle$ integrand $k^4 \frac{d}{d\tau}|A_{-\lambda}(\tau,\vec k)|^2$. Both curves are evaluated at $\tau=-1$, and displayed as a function of wavenumber, such that $\ln \left(\frac{k}{aH}\right) = 0$ corresponds to a horizon sized mode.  Left vertical line: wavenumber (or number of e-folds after horizon crossing) of the maximal exponential growth of $|A_{-\lambda}(\tau,\vec k)|^2$. Right vertical line: The $ \langle \vec E \vec B \rangle$ integrand gets its dominant contribution at about $\Delta N_\xi$ later. Here we have set $\xi = 5$. }
    \label{fig:WittakerEB}
\end{figure}

We shall now provide arguments that once $\xi$ becomes time-dependent, a second time scale (besides $H^{-1}$) appears, characterizing a resonance phenomenon with a frequency in e-fold time of $\omega_N^{res}=2\pi/\Delta N_\xi$. This resonance drives self-excited oscillations with frequency $\omega_N^{res}$ appearing in  $ \langle \vec E \vec B \rangle$.\\

Let us start our analysis by looking again at the gauge field Fourier mode equation of motion~\eqref{eq:eom_A}. Rewriting this into e-fold time
\be
dN=aH d\tau \quad\Rightarrow\quad \frac{d^2}{d\tau^2}=a^2 H^2 \left(\frac{d^2}{dN^2}+{(1-\epsilon)}\frac{d}{dN}\right) \quad ,
\ee
we get
\be
A_\pm''(\vec k)+{(1-\epsilon)}A_\pm'(\vec k) + \frac{k}{aH}\left(\frac{k}{aH} \pm 2 \lambda \xi \right) A_\pm(\vec k) = 0  \quad . \label{eq:eom_AN}
\ee
In the remainder of this section, we will neglect all terms suppressed by the slow-roll parameter $\epsilon = - H'/H \ll 1$ . {In our numerical analysis, described in Sec. \ref{sec:numerics}, we keep all slow-roll corrections though.}
We see that the mode $ A_{-\lambda}$ becomes tachyonic once $k/(aH)<2\xi$ , while it starts freezing out due to the friction term $ A_{-\lambda}'$ taking over once $k/(aH)<1/(2\xi)$. We now look at the behaviour of the mass term of the growing mode more closely. For constant $\xi$, the mass terms takes its maximally negative value $\hat{m}_{-\lambda}^2=-\xi^2$ at $k/(aH)=\xi$ since the quadratic function of $m^2_{-\lambda}=k/(aH)\left(k/(aH)- 2\xi\right)$ has zeroes at $k/(aH)=0$ and at  $k/(aH)= 2 \xi$. Hence, due the behaviour of the Whittaker function governing the gauge field modes, the major part of the growth of $A_{-\lambda }$ out of the Bunch-Davies initial conditions happens while $k/(aH)\simeq \xi$.

However, the integrand of $ \langle \vec E \vec B \rangle$, due to the $\tau$-derivative and the $k^4$ prefactor, takes its maximum contribution at approximately $k/(aH)=2/\xi$ (see also Appendix~\ref{app:phaseshift}). This implies that $ \langle \vec E \vec B \rangle$ is dominated by modes whose `knowledge' of the value of $\xi$ governing their maximum growth period originates from about 
\be
\Delta N_\xi\simeq \ln \frac{\xi^2}{2}
\label{eq:LagEstimate}
\ee
e-folds \emph{earlier}.
This is clearly visible in Fig.~\ref{fig:WittakerEB}, where we see that the $ \langle \vec E \vec B \rangle$ integrand $k^4\frac{d}{d\tau} |A_{-\lambda}(\tau,\vec k)|^2$ has its peak contribution about $\Delta N_\xi$ after the time when $|A_{-\lambda}(\tau,\vec k)|^2$ has its maximum exponential growth. Note, that in Fig.~\ref{fig:WittakerEB} we {took $\tau = -1$ and} 
expressed the wavenumber $k$ as number of e-folds after horizon crossing $-\ln k/aH$. {This means that the gauge modes are still sub-horizon at the time of maximal growth ($k/aH = \xi > 1$), but already super-horizon when they provide the peak contribution to the $ \langle \vec E \vec B \rangle$ integrand ($k/aH  = 2/\xi < 1$).}\\

Using this information, we can ask a simple question -- how does $ \langle \vec E \vec B \rangle$ react if we allow for a sudden step-like change of $\xi$ at a certain moment of time? For explicitness, let us assume that $\xi=\xi_0$ changes to $\xi_0+\Delta\xi>\xi_0$ at $N=N_0$ suddenly. At $N=N_0$ the integral $ \langle \vec E \vec B \rangle$ gets its dominant contribution from modes  $A_{-\lambda}(\vec k)$ with  $k/(aH) \simeq 2/\xi$ which had their growth happening $\Delta N_\xi$ e-folds earlier. At that time $N_0-\Delta N_\xi$ we still had $\xi=\xi_0$ and hence 
\be
| \langle \vec E \vec B \rangle_{N_0}| \simeq 2.4 \cdot 10^{-4}  H^4 \frac{e^{2 \pi \xi_0}}{\xi_0^4}\quad.
\ee
Conversely, modes  $A_{-\lambda}(\vec k)$ with $k/(a H) \simeq 2/\xi$ at $N=N_0$ will grow towards their plateau value and thus dominate $ \langle \vec E \vec B \rangle$ only starting at time $N=N_0+\Delta N_\xi$. These modes experience their growth for $N>N_0$ when $\xi>\xi_0$. Hence, they will approach a plateau governed by $\xi=\xi_0+\Delta\xi$ and thus
\be
|\langle \vec E \vec B \rangle_{N_0+\Delta N_\xi}| \simeq  2.4 \cdot 10^{-4}  H^4 \frac{e^{2 \pi (\xi_0+\Delta\xi)}}{(\xi_0+\Delta\xi)^4}> | \langle \vec E \vec B \rangle_{N_0}| \quad.
\ee

\begin{figure}[t!]
    \centering
    \includegraphics[width=0.6\textwidth]{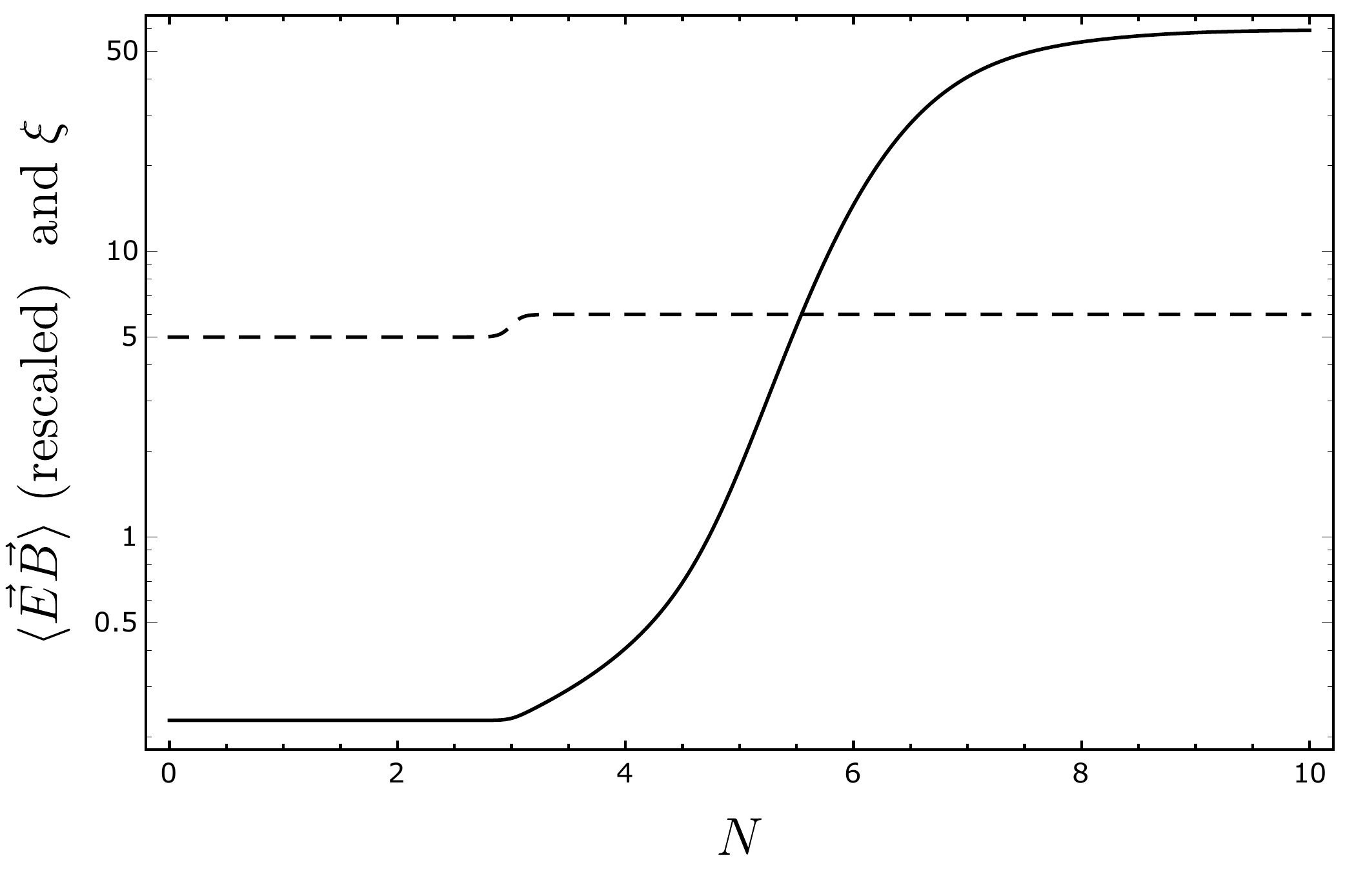}
    \caption{Black solid: Numerically computed, and rescaled, response of $\langle \vec E \vec B \rangle$ to the change in $\xi$ with significant lag  $\simeq \Delta N_\xi$. Black dash: Almost step function like change of $\xi$ modeled as $\xi(N)=\xi_0+\frac{\Delta\xi}{2}\left(1+\tanh(\mu_\xi (N-N_0))\right)$ with the jump taking place at {$N_0=3$} from $\xi_0=5$ with amplitude $\Delta\xi=1$ and steepness $\mu_\xi=10$  (dashed black).}
    \label{fig:StepFunctionLag}
\end{figure}

The transition from the initial plateau to the final plateau happens smoothly, yet clearly the system shows `lag': It reacts to a sudden change in $\xi$ by changing to its new  $\langle \vec E \vec B \rangle$ value only with a time lag of about $\Delta N_\xi$. A numerical computation of $\langle \vec E \vec B \rangle$ displayed in Fig.~\ref{fig:StepFunctionLag} clearly confirms this lag.\\

Assume now that instead of a sudden change, we provide $\xi$ with a periodic time dependence $\xi(N+2\pi/\omega_N)=\xi(N)$ with constant frequency $\omega_N$ in e-fold time. Clearly,  $\langle \vec E \vec B \rangle$ will now react with the same lag and thus oscillate with a phase shift
\be
\Delta\alpha=\omega_N\Delta N_\xi
\ee
as long as this phase shift $\Delta\alpha<2\pi$.\footnote{To see this from the `sudden approximation' argument before, break up a periodic $\xi(N)$ into small  step-wise changes.} Clearly then, demanding $\Delta\alpha=\pi$ as a necessary condition for resonance {(which can only occur if $\langle \vec E \vec B \rangle$ couples back to $\dot \phi$, this we will discuss shortly)}, this defines a critical frequency
\be
\omega_N^\star=\frac{\pi}{\Delta N_\xi}\quad.
\label{eq:critical_frequency}
\ee
\begin{figure}[t!]
    \centering
  \includegraphics[width=0.495\textwidth]{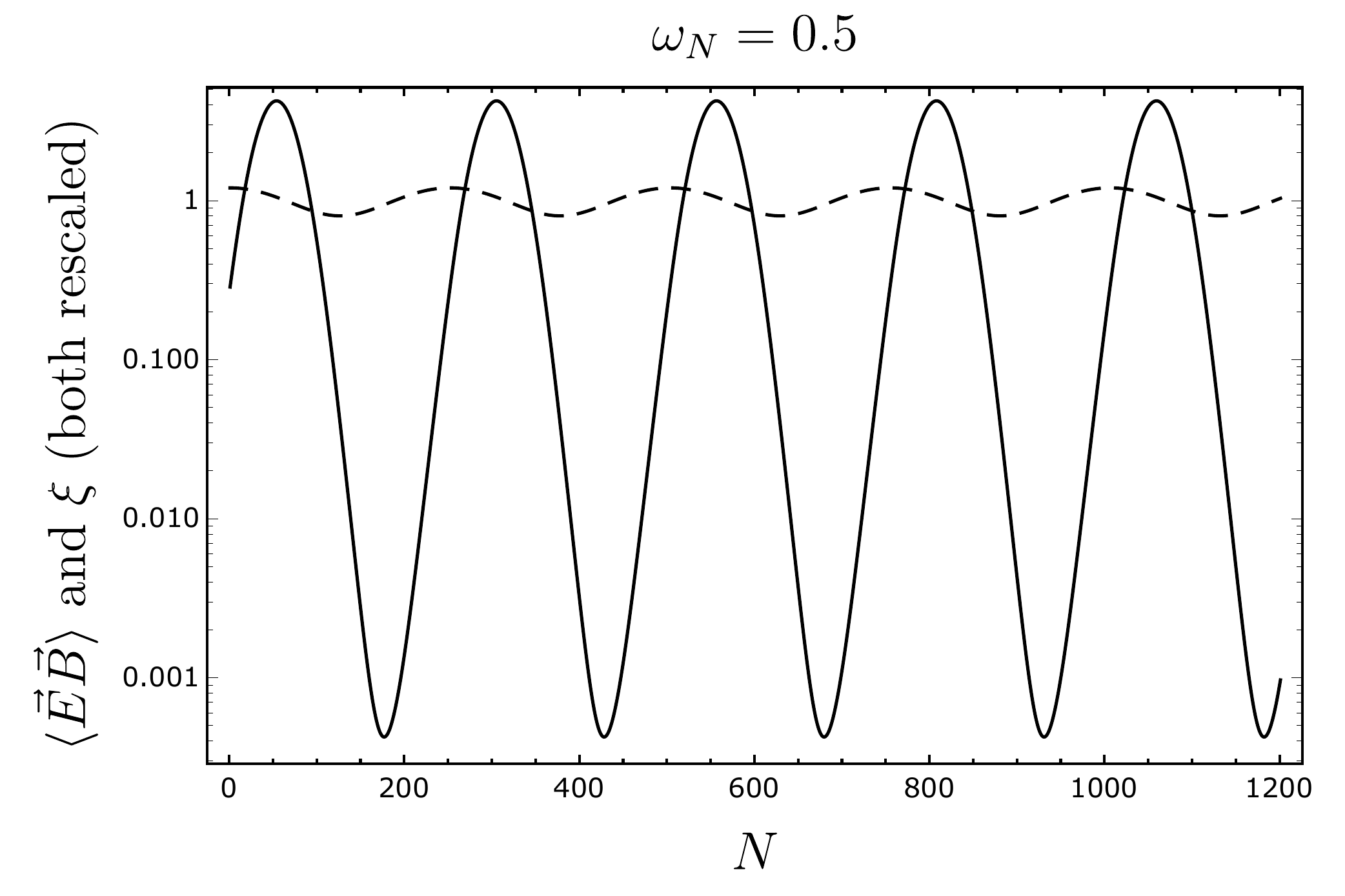}\hspace{3pt}\includegraphics[width=0.495\textwidth]{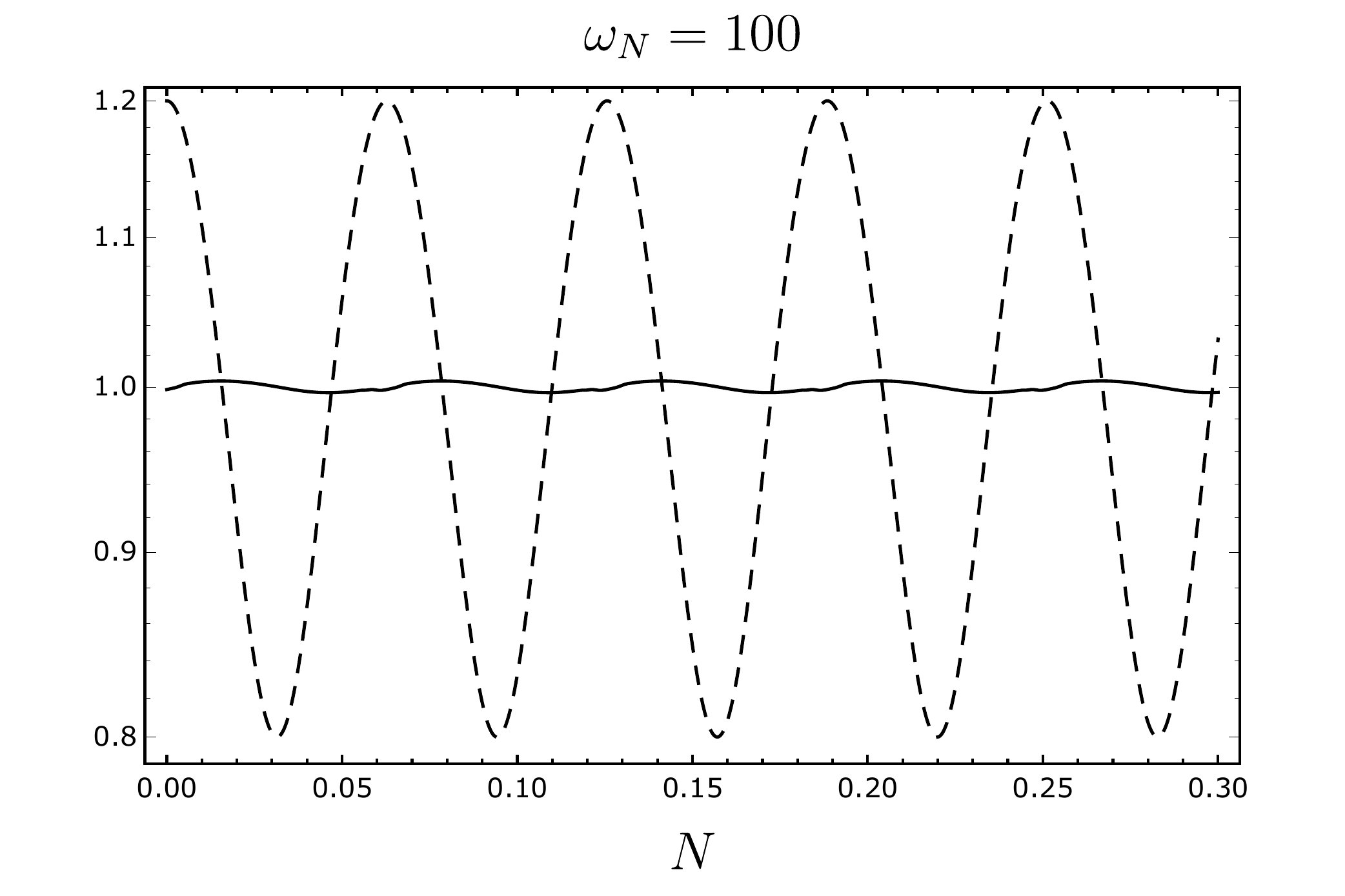}
    \caption{Left: Numerically computed, and rescaled, response of $\langle \vec E \vec B \rangle$ with significant lag (solid black) for a harmonic perturbation of $\xi$ with near-critical frequency $\omega_N\sim \omega_N^\star$ (dashed black). Right: For much larger frequencies the response averages out to zero. We chose $\bar\xi=5$ and the oscillation amplitude $\Delta\xi=1$.}
    \label{fig:EBvsNharmonicXi}
\end{figure}

We can numerically compute the full $\langle \vec E \vec B \rangle$ responding to a harmonic perturbation of $\xi$  around $\bar \xi$ with frequency $\omega_N$. Figure~\ref{fig:EBvsNharmonicXi} shows this for a frequency near $\omega_N^\star$, and for a frequency much larger than $\omega_N^\star$. We see clearly, that at $\omega_N\sim \omega_N^\star$ there is strong response of $\langle \vec E \vec B \rangle$ with lag. Moreover, at $\omega_N\sim \omega_N^\star$ the lag corresponds to a significant phase shift, while for much larger frequencies the response averages out to zero. 

\begin{figure}[t]
    \centering
    \includegraphics[width=0.6\textwidth]{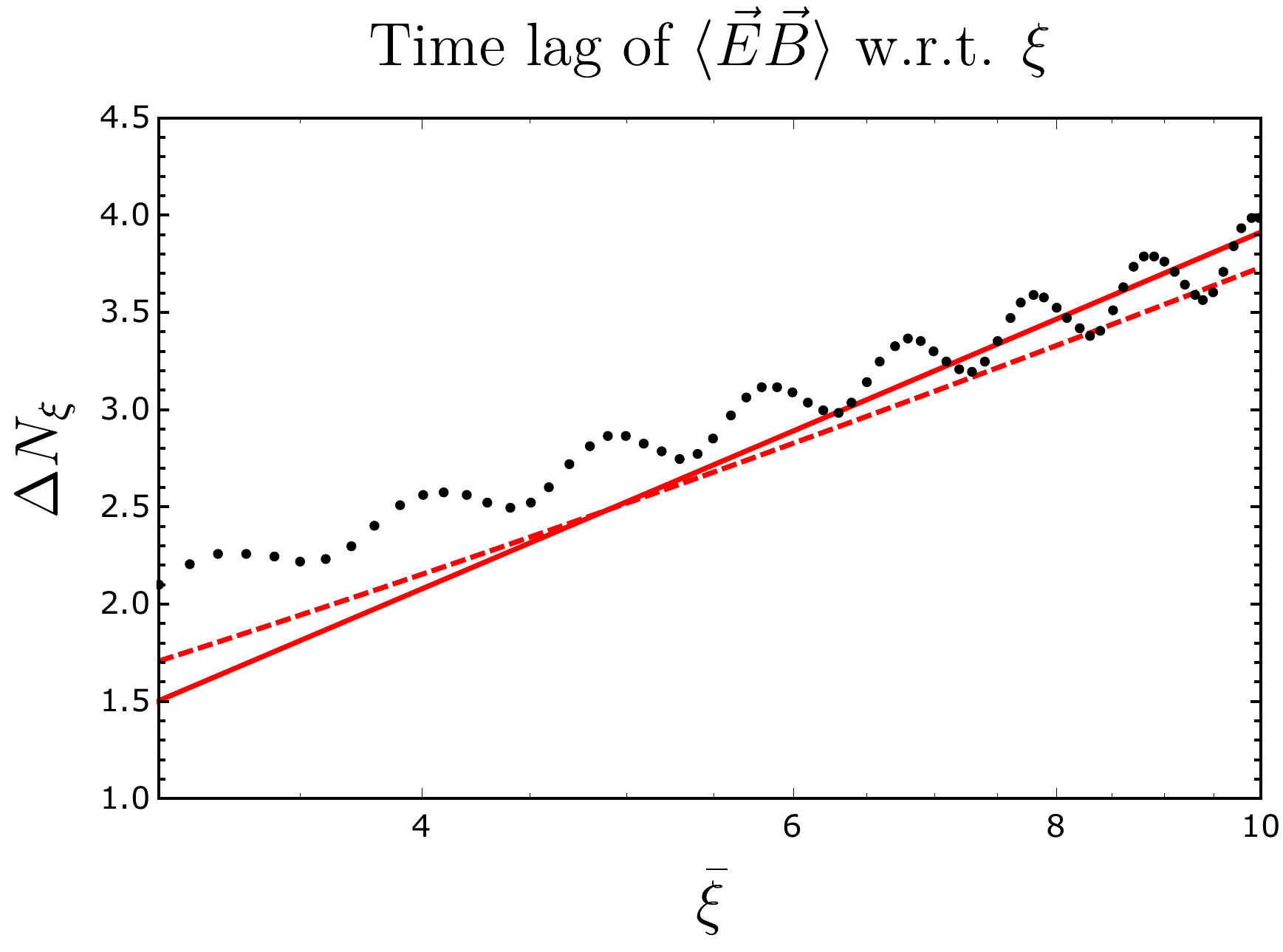}
    \caption{Data points: The lag $\Delta N_\xi$ for the numerically computed response of $\langle \vec E \vec B \rangle$ to a harmonic perturbation of $\xi$ with frequency $\omega_N=0.2$ as a function of $\xi$. Solid red line: our estimate $\Delta N_\xi\sim\ln(\xi^2/2)$ in Eq.~\eqref{eq:LagEstimate}. Dashed red line: refined estimate derived in App.~\ref{app:phaseshift}.}
    \label{fig:DeltaNvsXi}
\end{figure}

Finally, we can numerically determine the lag $\Delta N_\xi$ 
occurring as a function of $\xi$. This is shown in Fig.~\ref{fig:DeltaNvsXi} for $\omega_N = 0.2$ and clearly shows (solid red line) the scaling $\Delta N_\xi = \ln(\xi^2/2)$ derived in Eq.~\eqref{eq:LagEstimate}. The refined estimate derived in App.~\ref{app:phaseshift} is depicted by the dashed red line. 
{The oscillations visible at larger values of $\xi$ are not captured by the estimate~\eqref{eq:LagEstimate}, which was based on determining the difference between the points of maximal growth and maximal contribution to $\langle \vec E \vec B \rangle$ for any given mode at constant $\xi$. For a periodically varying $\xi$ these estimates receive corrections, which depend in particular on the shape of the pulses in the periodic function $\xi$.}\\

\begin{figure}[t]
    \centering
    \includegraphics[width=0.6\textwidth]{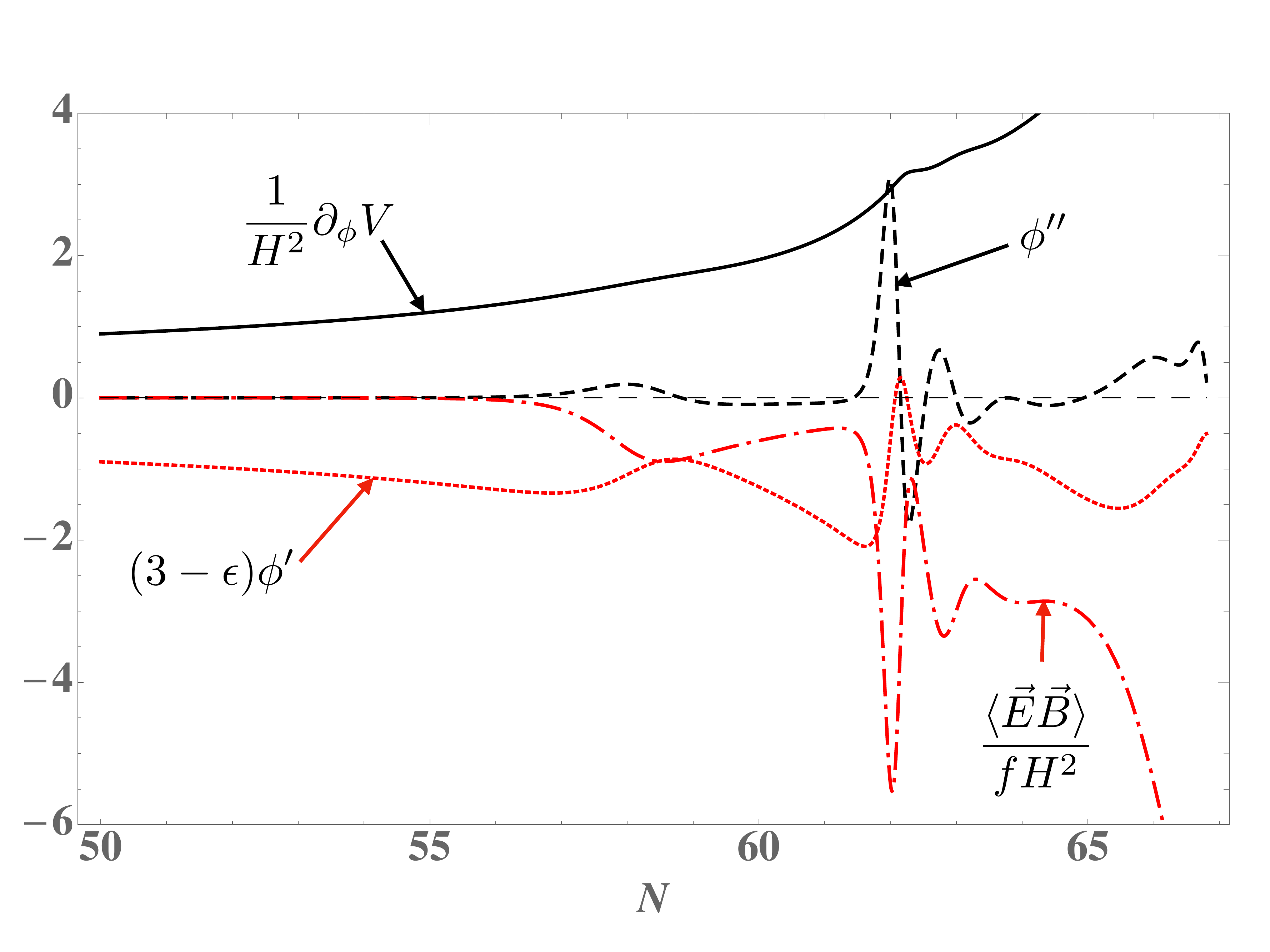}
    \caption{The contributions $\phi''$ (black dash), $\partial_\phi V/H^2$ (black solid), $(3-\epsilon)\phi'$ (red short dash), and $\langle \vec E\vec B\rangle/(fH^2)$ (red dash-dot) to the scalar field equation of motion {for $f = 1/25$ and $V(\phi) = m^2 \phi^2/2$ (see Sec.~\ref{sec:numerics})} [in units of $M_P$]. We have conveniently expressed the derivatives $\dot\phi$ and $\ddot\phi$ in terms of e-fold time derivatives $\phi',\phi''$. Note that for $N\lesssim 60$ we find that $\phi''$ is negligible, while the first long-wave oscillation has $\phi'$ and $\langle \vec E\vec B\rangle$ of opposite phase. Note further, that for $N\gtrsim 60$ the long-wave oscillations are superimposed by faster damped oscillations. For these, $\phi''$ is no longer negligible, and the phase shift at each step of the chain $\phi''\to\phi'\to\langle \vec E\vec B\rangle$ is about $\pi/2$.}
    \label{fig:phi_eom_contribs}
\end{figure}

At this point it becomes interesting to turn to our dynamically coupled system, where the $\xi$-parameter is determined by the scalar field equation of motion
\be
\ddot\phi+3H\dot\phi+V_{,\phi}-\frac{1}{f} \langle \vec E \vec B \rangle=0\quad.
\ee
The driving force of the scalar potential $V_{,\phi}$ is balanced by the sum of the Hubble friction (second term) and the gauge-field induced friction (contained in the last term), while the $\ddot \phi$ only becomes relevant in the very last stages of inflation.
 In our full numerical solution which clearly displays a resonance (see Sec.~\ref{sec:numerics}) we can observe that the  oscillating parts  of the two friction terms $3H\dot\phi$ and $ \langle \vec E \vec B \rangle$ (sourced by the time-dependent part of $\xi$) cancel against each other at $N\lesssim 60$ where the backreaction is not yet very strong, whereas $V_{,\phi}$, which depends only on $\phi$ but not on $\dot \phi$, evolves to good approximation monotonously. This is clearly visible in Fig.~\ref{fig:phi_eom_contribs} where we plot the different parts of the scalar field equation of motion evaluated on the numerical solution for $1/f=25$, discussed in detail in Sec.~\ref{sec:numerics}.

We now parametrize $\xi$ as $\xi =\bar\xi+\Delta\xi(N)$ with the long-time average $\overline{\Delta\xi}(N)\equiv\frac{1}{N} \int dN \Delta\xi(N)=0$, where an over-bar denotes averaging over time while all quantities are implicitly containing an average over separate universes part of the $\delta N$ formalism (unless this average is written explicitly as $\langle\ldots\rangle$).  Consequently, we can recast the time dependent part of $\dot \phi$ as $\Delta\xi(N)$ and get approximately
\be
6 H^2 \frac{f}{\lambda}\,\Delta\xi-\frac{1}{f} \Delta\langle \vec E \vec B \rangle(\Delta\xi)\simeq 0
\label{eq:EOMphiOscPart}
\ee
where $  \langle \vec E \vec B \rangle = \overline{\langle \vec E \vec B \rangle}+\Delta \langle \vec E \vec B \rangle$.

Now we use the properties of the background $ \overline{\langle \vec E \vec B \rangle}$ given in Eqs.~\eqref{eq:EB},\eqref{eq::AkWhit} to write 
\be
\overline{\langle \vec E \vec B \rangle}=-\lambda {\cal A}_{EB}
\ee
where ${\cal A}_{EB}>0$ is a positive definite function. Assuming the oscillating part $\Delta \langle \vec E \vec B \rangle$ will not change the sign of the total $\langle\vec E\vec B\rangle$, we can then define the split of $\langle\vec E\vec B\rangle$ into background and oscillatory part with a definite phase relative to the sign of $\overline{\langle \vec E \vec B \rangle}$ by writing
\be
\langle \vec E \vec B \rangle = \overline{\langle \vec E \vec B \rangle}+\Delta \langle \vec E \vec B \rangle\equiv-\lambda\left({\cal A}_{EB}+\Delta{\cal A}_{EB}\right)\quad.
\ee
This allows us rewrite Eq.~\eqref{eq:EOMphiOscPart} as
\be
\Delta\xi+\frac{1}{6f^2H^2}    \Delta{\cal A}_{EB}(\Delta\xi)=0\quad\Leftrightarrow\quad \Delta\xi=-\frac16 \,
\frac{\Delta{\cal A}_{EB}(\Delta\xi)}{f^2H^2}\quad.
\label{eq:dxi_eom}
\ee
Moreover, from the values of $f$ and $H$ we see that the factor $1/(6f^2H^2)$ rescales $\Delta{\cal A}_{EB}$ to be dimensionless and to have the same magnitude as $\Delta\xi$.

For this rescaled $\Delta{\cal A}_{EB}$, 
the discussion around Eq.~\eqref{eq:critical_frequency} and the numerical observation of the time delay in Fig.~\ref{fig:EBvsNharmonicXi} indicate the presence of a resonance at $\omega_N=\omega_N^\star$. The argument for this goes as follows: At the resonance frequency the observed time delay corresponds to a phase shift of $\phi$, that is, we observe
\be
\frac{ \Delta{\cal A}_{EB}(\Delta\xi(N))}{6f^2H^2}   \sim \Delta\xi\left(N-\frac{\pi}{\omega_N^\star}\right)\quad.
\label{eq:phase_shift_response}
\ee
Moreover, if we assume a nearly harmonic perturbation with an approximately constant frequency for $\Delta\xi$, we have by definition 
\be
 \Delta\xi\left(N-\frac{\pi}{\omega_N^\star}\right)\sim \Delta\xi''\quad.
 \label{eq:harmonic_relation}
\ee
Therefore, in plugging eq.~\eqref{eq:harmonic_relation} into eq.~\eqref{eq:phase_shift_response}, and this in turn into the right-hand side of Eq.~\eqref{eq:dxi_eom} we find that on a harmonic perturbation the equation of motion of $\xi$ becomes consistent with an oscillator equation.
\be
\Delta\xi\sim -\Delta\xi''\quad.
\ee 

Next, we observe that for  $N\gtrsim 60$ in Fig.~\ref{fig:phi_eom_contribs} there is a secondary pattern of damped oscillations at higher frequency compared to the long-wave 'base frequency' oscillations discussed above. For this pattern the oscillating contribution of $\ddot\phi$ is no longer negligible. Moreover, we observe that the phase shift at each step of the chain $\phi''\to\phi'\to\langle \vec E\vec B\rangle$ is about $\pi/2$. This implies that for this pattern the corresponding high-frequency (labeled by `h.f.') oscillating parts $\Delta\xi^{(h.f.)}$ and $\Delta{\cal A}_{EB}^{(h.f.)}$, split off the full quantities the same way as we did for the base frequency parts above, satisfy
\be
(\Delta\xi^{(h.f.)})'+3\Delta\xi^{(h.f.)}+\frac{1}{2f^2H^2}    \Delta{\cal A}_{EB}^{(h.f.)}(\Delta\xi)=0\quad.
\label{eq:dxi_eom_hf}
\ee
The observed phase relation in Fig.~\ref{fig:phi_eom_contribs} then states that $\Delta{\cal A}_{EB}^{(h.f.)}(\Delta\xi)$ has a phase shift of $\pi/2$ to the right compared to $\Delta\xi^{(h.f.)}$ and of $\pi$ to the right compared to $(\Delta\xi^{(h.f.)})'$. Hence, the figure indicates that for the high-frequency oscillations
\be
\Delta\xi^{(h.f.)}\sim (\Delta{\cal A}_{EB}^{(h.f.)})'\quad,\quad (\Delta\xi^{(h.f.)})'\sim (\Delta{\cal A}_{EB}^{(h.f.)})''\quad.
\ee
Plugging this relation into Eq.~\eqref{eq:dxi_eom_hf} we get the structure of the dampened harmonic oscillator differential equation
{\be
(\Delta{\cal A}_{EB}^{(h.f.)})''+{\cal O}(1) (\Delta{\cal A}_{EB}^{(h.f.)})'+(\omega^{(h.f.)})^2\Delta{\cal A}_{EB}^{(h.f.)}=0\quad.
\ee}
While we cannot determine the frequency of these faster oscillations $\omega^{(h.f.)}$ at this time, we consider the fact that the equation of motion takes the dampened oscillator form to be strong evidence supporting the existence of these secondary, faster dampened oscillations in the coupled system.

It is due to this line of reasoning that we conclude the presence of resonance occurring in the strong gauge-field back-reaction regime. Neglecting the resonance phenomenon, $\xi$ is typically a monotonically growing function of $N$, while the resonance frequency only scales logarithmically with $\xi$ and thus $N$. Hence, the sweep of $\xi$ effectively scans over possible resonance frequencies. Hence we expect the increasing value of $\xi$ to eventually trigger the resonance behaviour with approximately the predicted frequency. Some of the ideas presented here have been qualitatively previously presented in Refs.~\cite{Cheng:2015oqa,Notari:2016npn,DallAgata:2019yrr}. After formalizing these arguments, we here succeed in quantitatively explaining the observed resonance frequency.
Strictly speaking, the arguments spelled out above form a necessary, but not sufficient condition to ensure a resonance. However, in our numerical solutions to this coupled system of differential equations (see next section) we always see this resonance, indicating that this is indeed a generic feature.

\section{Numerical results \label{sec:numerics}}
We performed a full numerical analysis taking $M_P/f = \{20,25\}$ and  $V(\phi) = m^2 \phi^2/2$ with $m=6\times 10^{-6}\, M_P$, reproducing the observed amplitude of the scalar power spectrum at CMB scales.\footnote{As expected for the discussion in Sec.~\ref{sec:toymodel}, the generic features of the results discussed here are not very sensitive to the precise form of the scalar potential. In particular, we confirm similar results using a potential linear in $\phi$.}
Our final goal is to find the solution of the system of coupled integro-differential equations (\ref{eq:eom_phi_0}), (\ref{eq:eom_A}) and (\ref{eq:EB}).  The first step is to solve the inflaton equation of motion using the estimate of $\langle \vec{E}\vec{B}\rangle$ given in Eq.~(\ref{eq:EB_estimate}), which is obtained by solving the equations of motion of the gauge field modes, $A_{-\lambda}(\tau,k)$, assuming a constant inflaton speed, Eq.~(\ref{eq::AkWhit}). Then, choosing an appropriate array of $k$-modes, we solve Eq.~(\ref{eq:eom_A}) for each mode and we compute the discretized integral of equation Eq.~(\ref{eq:EB}), getting a new estimate of the backreaction.
We reach the final solution by iterating this procedure until we reach the end of inflation with a self-consistent solution, see App.~\ref{app:numerics} for details.
The initial conditions for the inflaton field are chosen at CMB scales in accordance with the vacuum slow-roll solution
while the $A_k$ modes satisfy Bunch-Davies vacuum conditions; we stop the time evolution when the system reaches the end of inflation $\epsilon\simeq1$. 

\begin{figure}[h]
    \centering
    \includegraphics[width=0.505\textwidth]{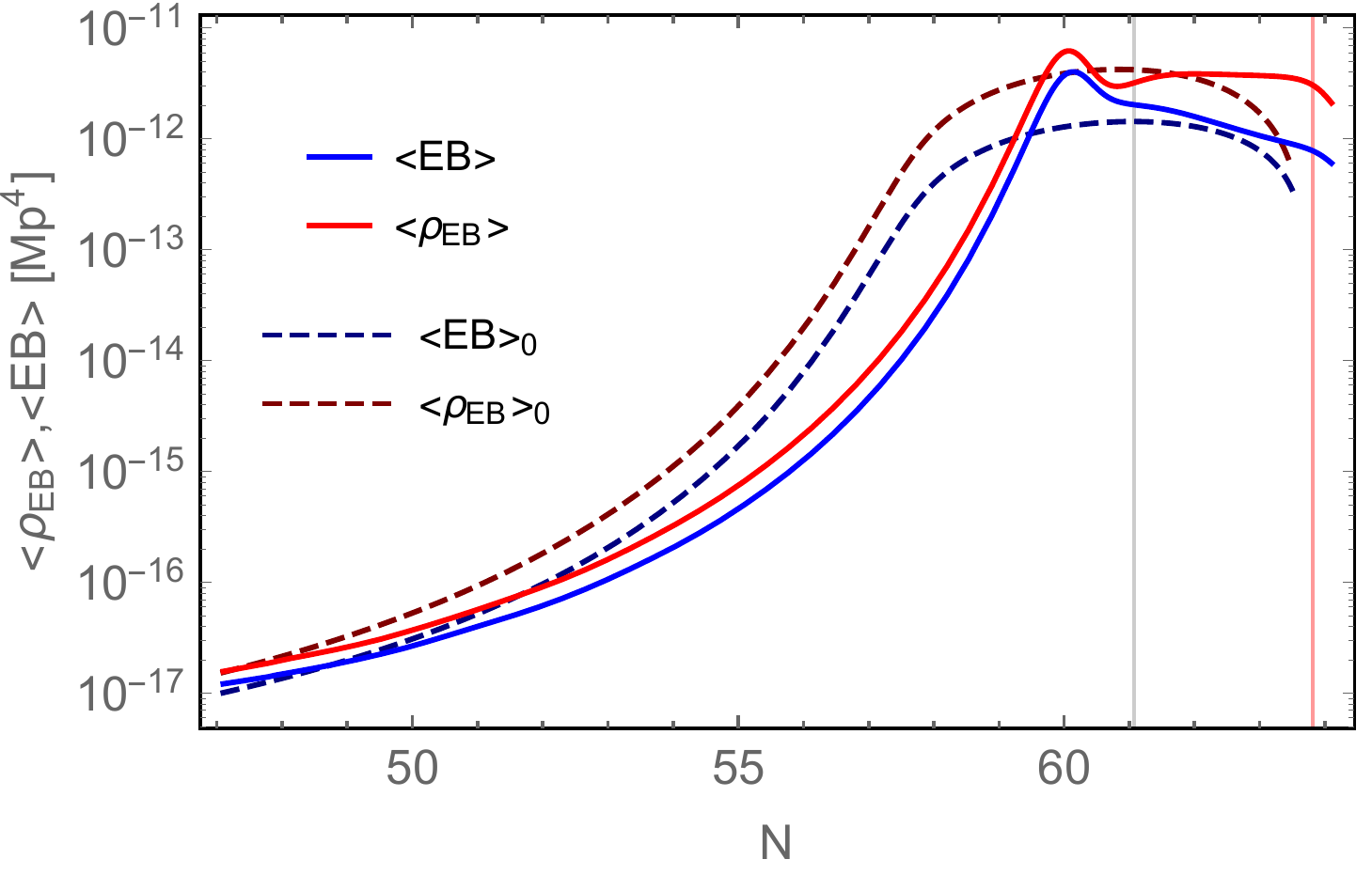}\hspace{3pt}\includegraphics[width=0.47\textwidth]{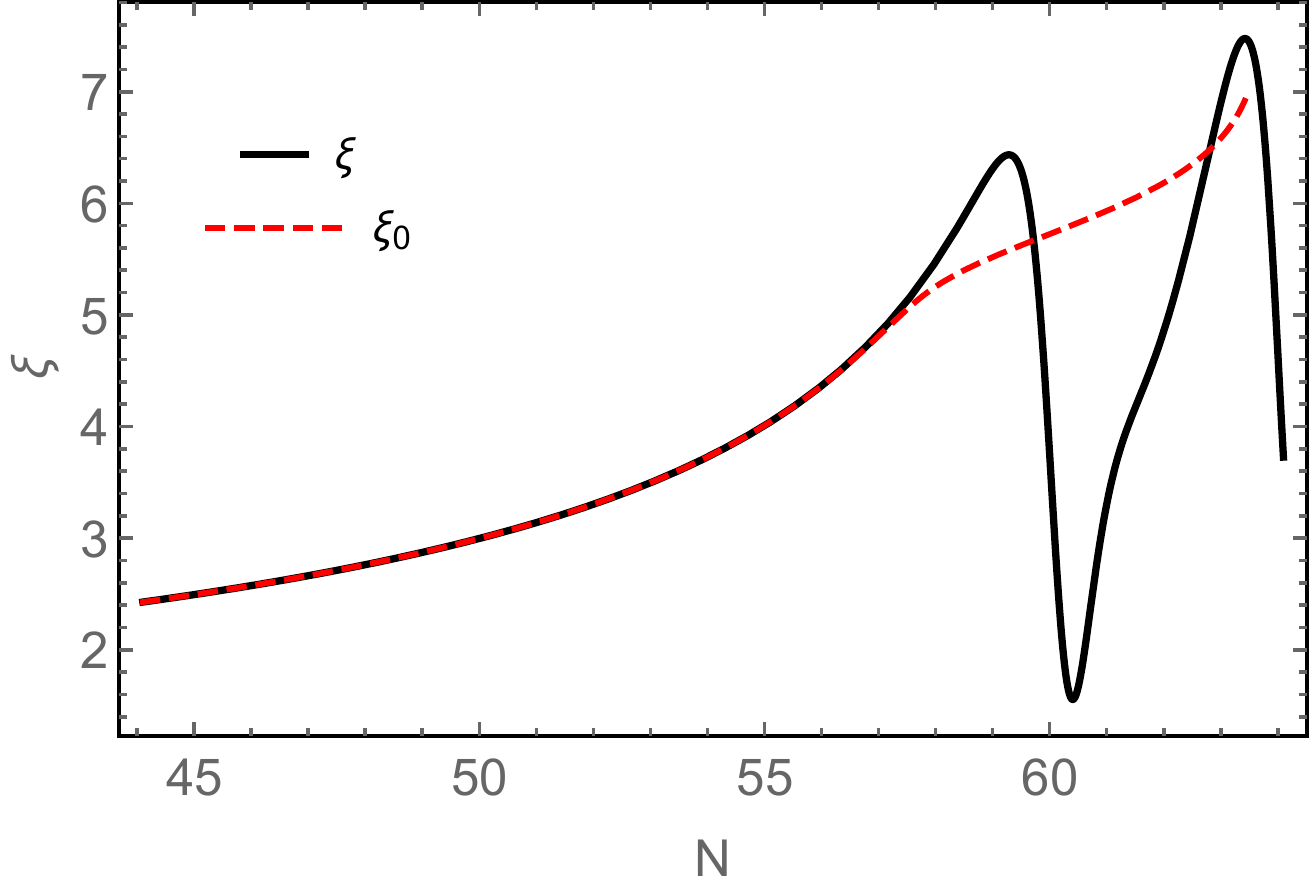}
    \includegraphics[width=0.503\textwidth]{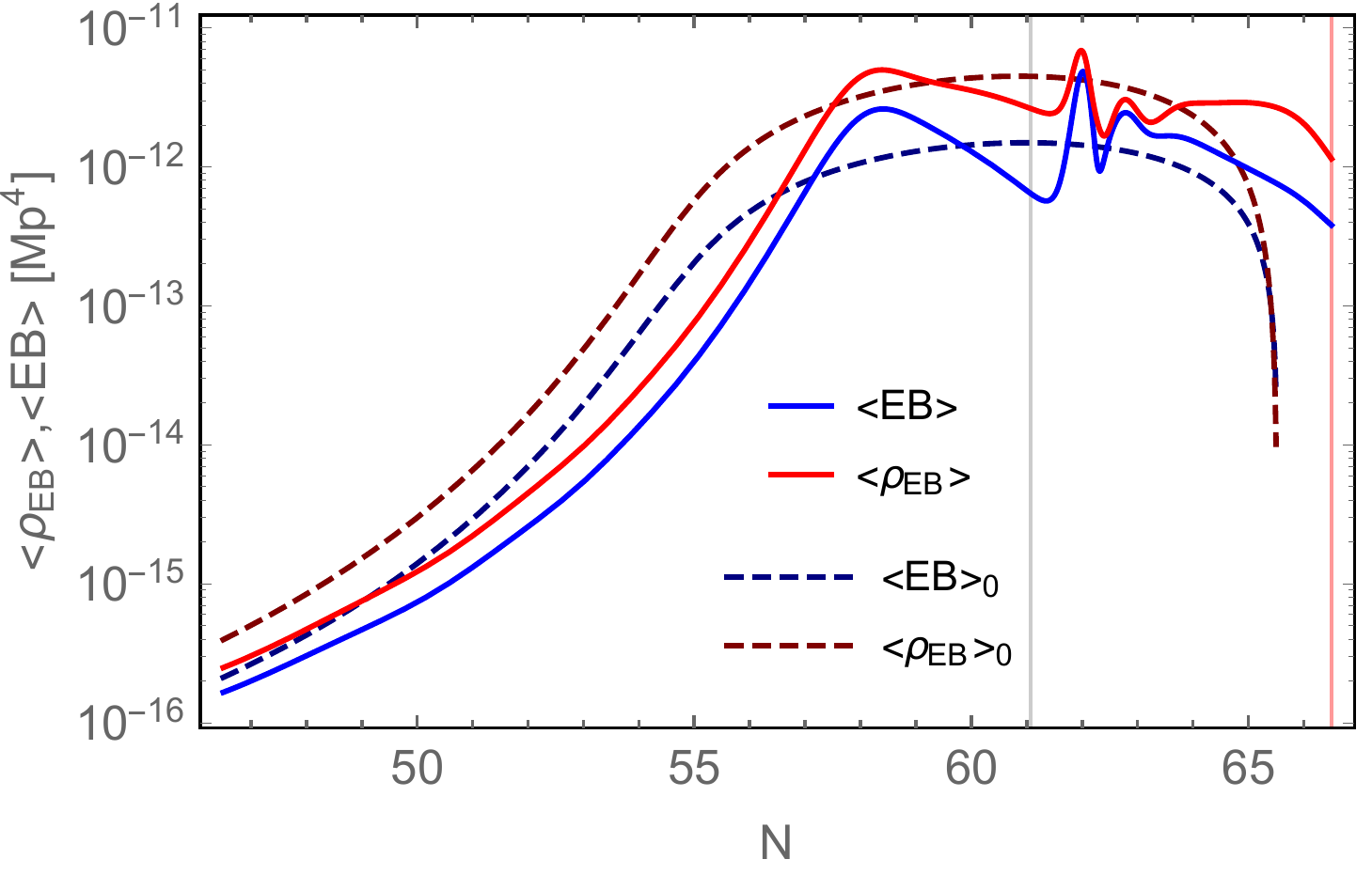}\hspace{3pt}\includegraphics[width=0.472\textwidth]{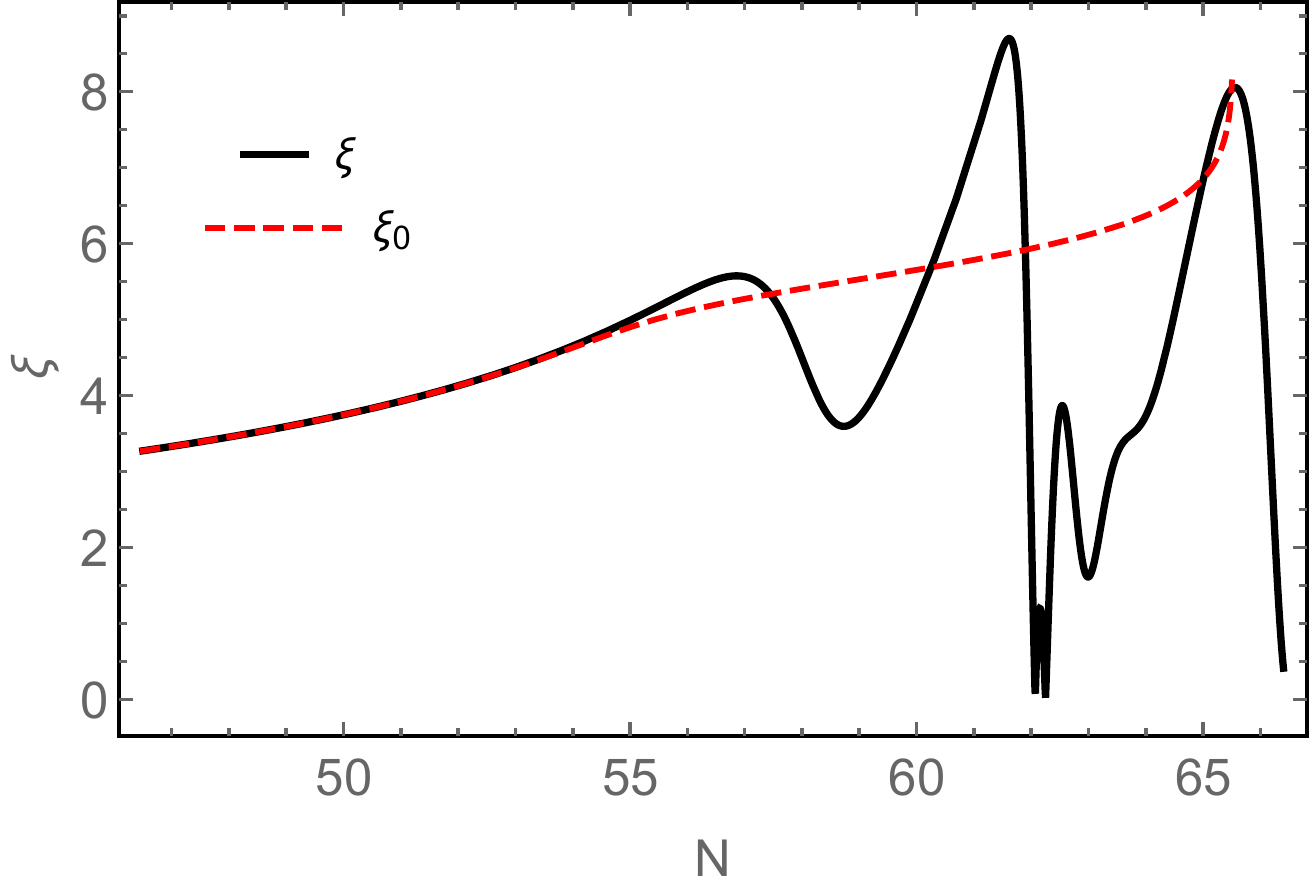}
    \caption{Top: $1/f=20$. Bottom: $1/f=25$. The left panels show the numerical results of $\rho_{EB}$ and $\langle \vec{E}\vec{B}\rangle$ (solid lines) compared to their analytical estimate (\ref{eq:EB_estimate}), (\ref{eq:EEBB_estimate}) (dashed lines). The vertical lines refer to the end of inflation in absence of backreaction (black line) and for the full numerical analysis (red line). The right panels show the oscillatory behaviour of the $\xi$ parameter (solid black line) compared to its analytical result coming from the solution of the inflaton equation of motion when the gauge field backreaction is given by (\ref{eq:EB_estimate}) (dashed red line). For better visibility, we display only the last $\sim20$ e-folds of inflation.
 }
    \label{fig:mesh1}
\end{figure}

The results of our analysis for $1/f=\{20,25\}$ are shown in Fig.~\ref{fig:mesh1} where we compare the final solution for $\langle \vec{E}\vec{B}\rangle$ and $\rho_{EB}=\langle \frac{E^2+B^2}{2}\rangle$ with the analytical estimate of Eqs.~(\ref{eq:EB_estimate}) and (\ref{eq:EEBB_estimate}). We also plot the $\xi$ parameter which shows that the oscillatory behaviour of the inflaton speed becomes more apparent in case of strong backreaction.\footnote{At the maxima of these oscillations, the value of $\xi$ exceeds the threshold  $\xi \simeq 4.7$ bounding the perturbative regime for approximately constant $\xi$~\cite{Ferreira:2015omg,Peloso:2016gqs}. This threshold cannot be immediately applied to a strongly oscillating $\xi$ and we will comment on perturbativity constraints in more detail in Sec.~\ref{sec:PS}.} We see that the numerical solution including the backreaction oscillates around the analytical estimate, with an oscillation period of $\Delta N_\xi \sim 3$, in accordance with our estimate in Sec.~\ref{sec:toymodel}. For $f = 1/25$ the value of  $\phi'$ temporarily changes sign (at $N \simeq 62$). The reason for this is the delay in gauge friction term discussed in Sec.~\ref{sec:toymodel}. As $|\phi'|$ drops, the gauge friction drops and the opposite sign of $\phi'$ (encoded by $\lambda$) entails the opposite sign for the gauge friction term as one would expect of a friction term. However, since the gauge friction term is dominated by modes which are controlled by the value of $\phi'$ some $\Delta N_\xi$ e-folds earlier, the sign change in the gauge friction term is delayed, allowing $\phi'$ to temporarily change sign. 

Our results are in accordance with those previously found in Refs.~\cite{Cheng:2015oqa,Notari:2016npn,DallAgata:2019yrr}, which reported oscillatory features in the inflaton velocity with a period of $3 - 5$ e-folds. All these studies are based on fully independent codes and numerical methods, and the results observed can be nicely explained with the semi-analytical arguments presented in Sec.~\ref{sec:toymodel}.

\section{Scalar power spectrum and primordial black holes \label{sec:PS}}

\subsection{Scalar power spectrum sourced by gauge field configuration}
The gauge field population does not only backreact on the dynamics of the homogeneous inflaton field but also acts as source term for the scalar inhomogeneities sourcing the density perturbations of the Universe. In the separate universe picture, curvature fluctuations on super-horizon scales are obtained as~\cite{Starobinsky:1986fxa,Salopek:1990jq,Sasaki:1995aw,Yokoyama:2007uu}~\footnote{This expression relies on the assumption that $\Delta N(\phi_1, \phi_2)$, the time in e-folds required for the inflaton to move from $\phi_1$ to $\phi_2$ does not depend on any further independent parameters, such as e.g.\ the inflaton velocity. For the attractor solution, this is justified even taking into account the strong, velocity-dependent friction. In the strongly oscillatory phase towards the end of inflation we expect corrections due to the break-down of the slow-roll approximation. \label{ft:deltaN} }
\begin{align}
 \zeta_c \simeq \delta N(t_*) \simeq N_{, \phi}(t_*) \; \delta \phi(t_*) \,.
 \label{eq:deltaN}
\end{align}
Here $N(t_*)$ denotes the average number of e-folds elapsed between $t_*$ and the end of inflation, whereas $\delta N(t_*)$ denotes the deviation occurring in a particular patch of the Universe induced by super-horizon scalar fluctuations. The perturbed version of Eq.~\eqref{eq:eom_phi_0} reads
\begin{align}
 0 = & \,  \phi'' + \left( 3 + \frac{H'}{H} \right) \phi' + \frac{V_{,\phi}}{H^2} \nonumber \\
 &  + \frac{H'}{H} \delta \phi' + \phi' \,  \frac{\partial}{\partial N} \left( \frac{H'}{H} \right) \delta N +   \frac{\partial}{\partial N} \left( \frac{V_{, \phi}}{H^2} \right) \delta N + \frac{2 H'}{f H^3} \langle \vec E \vec B \rangle \, \delta N  \nonumber \\
 &+ \delta \phi''  + 3 \, \delta \phi' - \frac{1}{f H^2} \vec E \vec B  - \frac{1}{f H^2} \frac{\partial \langle \vec E \vec B \rangle}{\partial N} \delta N \,. 
 \label{eq:deltaphi_0}
\end{align}
Since we are keeping only fluctuations to first order, all occurrences of $H$, $V$ and $\langle \vec E \vec B \rangle$ are here understood to be evaluated in terms of the homogeneous field $\phi$. On the contrary, the factor $\vec E \vec B$ in the third term of the third line includes the inhomogeneities in the gauge fields sourced by $\delta \phi$. Using Eq.~\eqref{eq:eom_phi_0} to replace the terms in the first line, dropping the slow-roll suppressed terms in the second line and inserting Eq.~\eqref{eq:deltaN} this simplifies to
\begin{align}
 L_{N}[\delta \phi(N)] \equiv \delta \phi'' + 3 \, \delta \phi'   - \frac{N_{,\phi}}{f H^2} \frac{\partial \langle \vec E \vec B \rangle}{\partial N} \delta \phi  =  \frac{1}{f H^2} (\vec E \vec B - \langle \vec E \vec B \rangle) \equiv \frac{1}{f H^2} \delta_{EB} \,.
 \label{eq:deltaphi_1}
\end{align}

This inhomogeneous linear differential equation can be solved by the Greens function method, see e.g.~\cite{Anber:2009ua,Barnaby:2011vw}.\footnote{For a comparison with these pioneering works see App.~\ref{app:comparison}. In short, we confirm the results found in the weak backreaction regime but disagree in the strong backreaction regime. We find the backreaction to be weaker than previously estimated, leading to a significant enhancement of the scalar power spectrum in this regime.}\\
For any linear operator $L_N$, the Greens function satisfying
\begin{align}
 L_N \, G(N, N') = \delta(N - N') \,,
 \label{eq:green}
\end{align}
can be convoluted with the source term $S(N)$,
\begin{align}
 \delta \phi(N) = \int G(N,N') S(N') dN' \,,
\end{align}
to obtain a solution of the inhomogeneous equation $L_N\, \delta \phi(N) = S(N)$. In Eq.~\eqref{eq:deltaphi_1} we identify $S(N) = \delta_{EB}/(f H^2)$. Moreover, for any given function $\langle \vec E \vec B \rangle(N)$ we can determine (at least numerically) the Greens function of the corresponding linear operator $L_N$ by solving the ordinary differential equation~\eqref{eq:green}. Since this is a second order differential equation we need to specify two boundary conditions which we take to be $G(N,N) = 0$ and $G'(N,N) = 1$.\footnote{{For the retarded Green's function $G(N,N')=0$ if $N'>N$. In addition we know that $G(N,N')$ must be a continuous function since $L_N G(N,N')$ does not involve generalized functions beyond $\delta(N-N')$ functions and in particular it does not contain derivatives of $\delta$ functions. Imposing continuity at equal time requires $\displaystyle\lim_{N'\rightarrow N_-} G(N,N')=\lim_{N'\rightarrow N_+}G(N,N')=0$.  On the other hand, integrating (\ref{eq:green}) over an infinitesimal neighbourhood of $N=N'$ we get $\displaystyle\int_{N'-\epsilon}^{N'+\epsilon}  L_N G(N,N') dN=1$. $G$ being  continuous, $\partial_N G$ must be bounded and we immediately see that if we shrink the integration domain to zero size the only term which can give a finite contribution is $\displaystyle\lim_{\epsilon\rightarrow 0}\displaystyle\int_{N'-\epsilon}^{N'+\epsilon}  L_N G(N,N') dN=\displaystyle\lim_{\epsilon\rightarrow 0}\displaystyle\int_{N'-\epsilon}^{N'+\epsilon}  \partial_{N}^2G(N,N') dN=\partial_N G(N'_+,N')- \partial_N G(N'_-,N')=\partial_N G(N'_+,N')=1$.}}

With this, the two-point function of scalar perturbations exiting the horizon at e-fold $N$ can be computed as
\begin{align}
 \langle \zeta^2 \rangle & = \langle \delta N^2 \rangle = N_{,\phi}^2 \langle \delta \phi^2 \rangle 
  = N_{,\phi}^2 \int dN' G(N,N') \int dN''  G(N, N'') \langle S(N') S(N'') \rangle\,.
  \label{eq:PS_masterformula}
\end{align}
{We parametrize the unequal time correlations by $g(N', \Delta N)$,
\begin{align}
  \int_{N' + \Delta N}^\infty dN'' \langle S(N') S(N'') \rangle  = \langle S(N')^2  \rangle  g(N', \Delta N)  \,.
  \label{eq:g_def}
\end{align}
with
\begin{align}
 g(N', \Delta N) = \begin{cases}
                    \gamma & \Delta N = 0 \\
                    \epsilon & \Delta N > 0
                   \end{cases}
\end{align}
where $\gamma = {\cal O}(1)$ and $\epsilon \rightarrow 0$ in the limit of vanishing unequal time correlators, i.e.\ in the limit of white noise. If $G(N,N'')$ and $\langle S^2(N') \rangle$ do not vary significantly over the support of $g(N', \Delta N)$ we can approximate\footnote{
{To verify these approximations and quantify the importance of the unequal time contributions, we numerically evaluate $g(N',\Delta N)$ using the mode functions $A_k(N)$ from the numerical computation in Sec.~\ref{sec:numerics}. See App.~\ref{app:NETcorr} for details. Far away from the resonance regime, we find this approximation to be unproblematic. As we approach the resonant regime, the unequal time correlators become more important while at the same time $\sigma^2_\text{EB}$ varies more rapidly. We find values of $g(N', 0.1)/\gamma \simeq 0.9$ and $g(N', 0.5)/\gamma \simeq 0.4$, indicating that most of the support of $g$ is focused on a small region over which $\sigma^2_\text{EB}$ varies only moderately. We conclude that the unequal time correlators most likely lead to an ${\cal O}(1)$ correction to \eqref{eq:powerspectrum_et} in the resonance regime, slightly smearing out the peaks and troughs.} 
}
\begin{align}
  \int dN'' G(N, N'') \langle S(N') S(N'') \rangle & \simeq G(N,N') \langle S^2(N') \rangle g(N',0) \nonumber \\
   & \simeq \frac{G(N,N')}{f^2 H^4} \langle \delta_{EB}^2(N') \rangle  \nonumber  \\
 & = \frac{G(N,N')}{f^2 H^4} \sigma^2_{EB}(N') \label{eq:delta_as} \,,
\end{align}}
with $\sigma_{EB}^2 \equiv (\vec E \vec B - \langle \vec E \vec B \rangle)^2$ denoting the variance of $\vec E \vec B$ at a given time. For a given set of mode functions $A_k(N)$ the variance $\sigma^2_{EB}$ can be computed explicitly, see e.g.\ App.~A of \cite{Linde:2012bt}. 
The final expression for the power spectrum then reads
\begin{align}
 \Delta_\zeta^2 =  \langle \delta \zeta^2 \rangle \simeq N_{,\phi}^2 \int dN' \frac{G^2(N,N') \sigma_{EB}^2(N')}{f^2 H^2(N')} + \langle \zeta^2 \rangle_\text{vac}\,,
 \label{eq:powerspectrum_et}
\end{align}
where $\langle \zeta^2 \rangle_\text{vac}^{1/2} = H/(2 \pi \phi')$ is the usual vacuum contribution.
\begin{figure}
\center
 \includegraphics[width = 0.49 \textwidth]{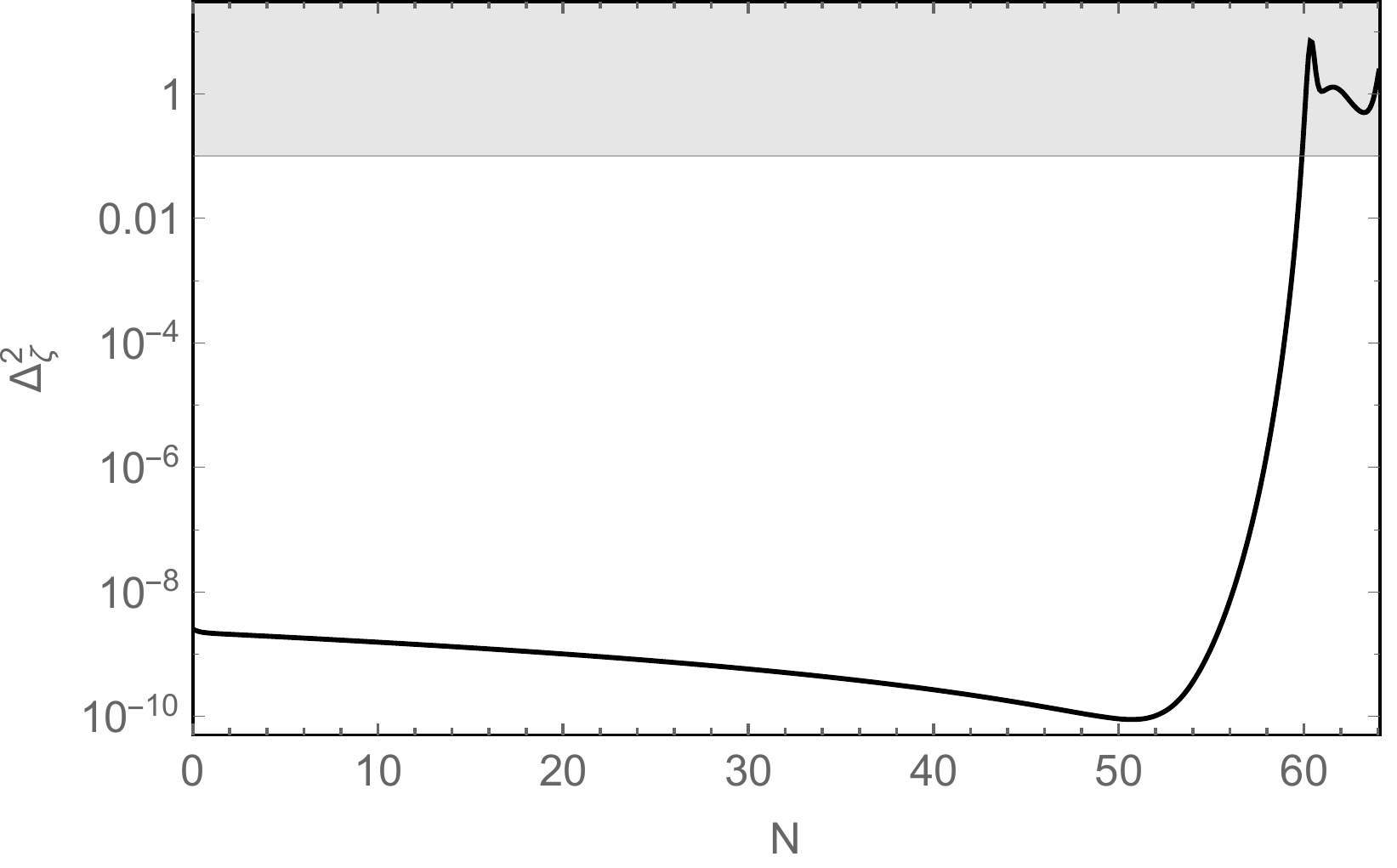} \includegraphics[width = 0.49 \textwidth]{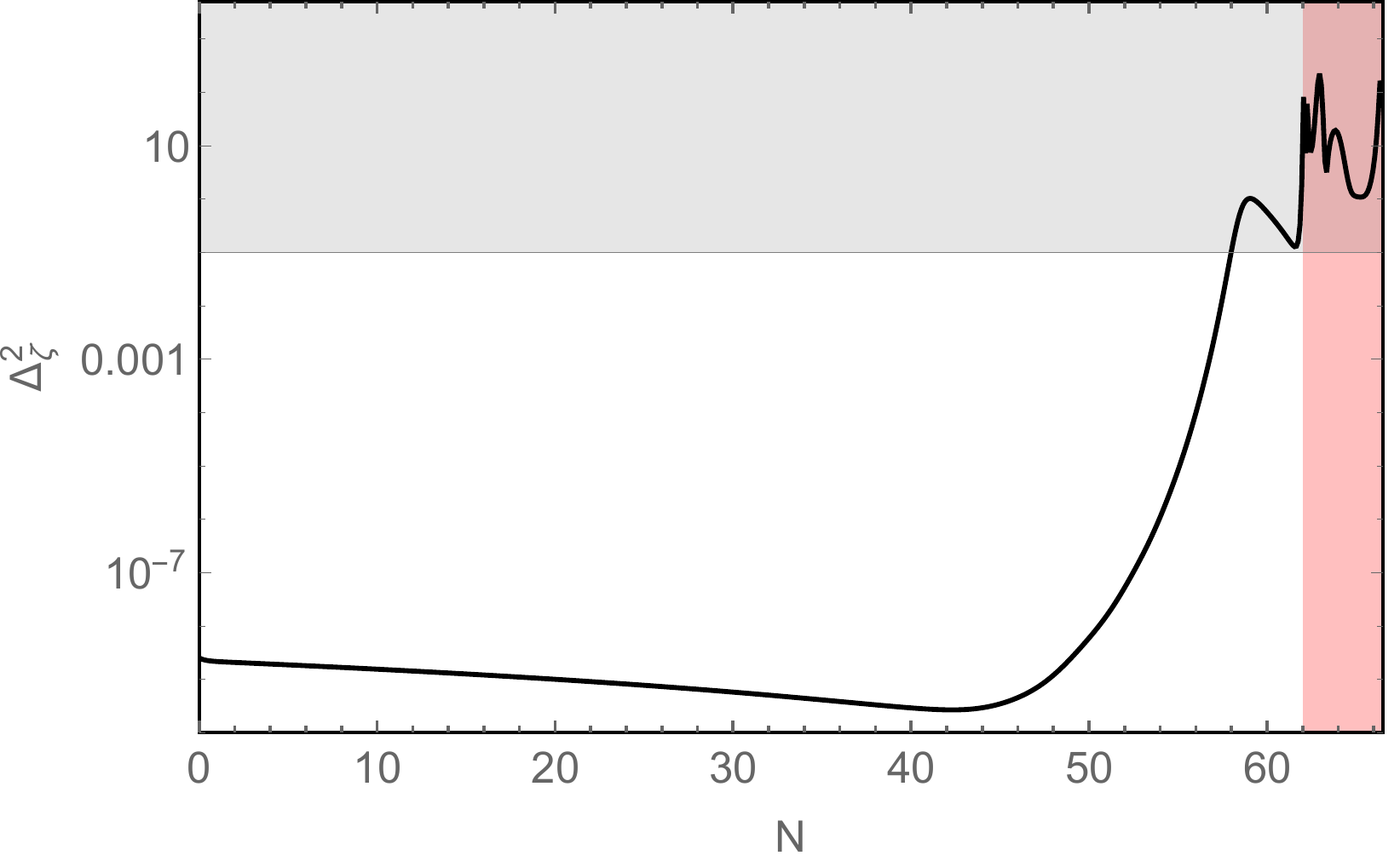}
 \caption{Scalar power spectrum for $1/f = 20$ (left) and $1/f = 25$ (right). The resonantly enhanced gauge field population leads to strong enhancement of the scalar power spectrum at small scales, with peaks reflecting the resonance structure. The gray and red shaded areas indicate the limitations of the $\delta N$ formalism, see text for details.}
 \label{fig:PSnum}
\end{figure}

The result obtained by numerically evaluating the Greens function $G(N,N')$ and the variance $\sigma_{EB}$ is depicted in Fig.~\ref{fig:PSnum}. The power spectrum is dramatically enhanced towards the end of inflation and inherits the resonant oscillations present in the source term. As highlighted by the gray band, the power spectrum extends above $\zeta \sim 0.3$, indicating the breakdown of the perturbative expansion used in our analysis. Moreover, for $f = 1/25$, the inflaton speed temporarily changes sign (see Fig.~\ref{fig:mesh1}), implying that $\phi$ is not monotonously increasing. Strictly speaking, this requires to go beyond the standard $\delta N$ formalism (see footnote~\ref{ft:deltaN}). In practice, since this only happens for a very short period of time, we expect the $\delta N$ formalism (with the inflaton speed regularized to some small value round $N \simeq 62$) to nevertheless give a good estimate. The corresponding problematic region is highlighted in red in the right panel of Fig.~\ref{fig:PSnum}. Due to these caveats, we cannot make a prediction about the precise amplitude of the scalar power spectrum at small scales. However, we can conclude that power spectrum reaches values of $\Delta_\zeta^2 \gtrsim 0.01$ in the last e-folds of inflation, exceeding the threshold for primordial black hole formation (see below).

The very large values for the scalar perturbations at small scales, indicating an inhomogeneous field configuration with large gradient energy, may trigger a premature end of inflation. This would relax the bounds from primordial black hole formation and consequently the bound on the coupling $1/f$ (see below). However, recent findings~\cite{Kleban:2016sqm,Clough:2016ymm,Clough:2017efm,Aurrekoetxea:2019fhr} indicate that high-scale inflation is quite robust against large gradient energies. How much of this stability against large gradients remains on the $\simeq 2..3 \, M_{\rm P}$  of field range corresponding to the last about 5 e-folds of inflation in a quadratic potential is an open question which we leave for future work. We hope that our findings will trigger a more detailed non-perturbative analysis of this last stage of inflation.

Even discarding the peaks arising from the resonant enhancement, the amplitude of the power spectrum in Fig.~\ref{fig:PSnum} at small scales is significantly larger than expected from previous estimates~\cite{Anber:2009ua,Linde:2012bt}. We provide a detailed comparison and discussion in Appendix~\ref{app:comparison}. In summary, we conclude that previous analytical analyses have overestimated the amount of backreaction in Eq.~\eqref{eq:deltaphi_1} and have hence underestimated the amplitude of the power spectrum in the strong backreaction regime. Consequently, the amplitude of the scalar power spectrum we report is in particular significantly larger than found in \cite{Cheng:2015oqa}, which accounted for the oscillating inflaton velocity but used the estimate for the power spectrum derived in~\cite{Linde:2012bt}.

\subsection{Primordial black hole formation and phenomenology}

If the scalar perturbations at a given scale exceed a critical threshold $\zeta_c \sim 0.5$ they collapse into a primordial black hole upon horizon re-entry~\cite{Carr:2009jm}. The mass of the corresponding black hole is determined by the energy contained in a Hubble volume at the time of horizon re-entry,
\begin{equation}
 M_{PBH}(N) \simeq \gamma \, \frac{4 \pi }{3} \, (e^{- j N} H_\text{inf})^{-3} \times 3 \,  (e^{- j N} H_\text{inf})^2 \,M_P^2 \simeq 55~\text{g} \, \gamma \left( \frac{10^{-6} M_P}{H_\text{inf}}\right) e^{jN}\,,
 \label{eq:PBHmass}
\end{equation}
with $N$ counting the number of e-folds from the horizon exit of the respective fluctuation until the end of inflation, $H_\text{inf}$ denoting the Hubble parameter at this time, $j = 2$ ($j = 3$) for radiation (matter) domination after inflation and $\gamma \simeq 0.4$ parametrizes the efficiency of the gravitational collapse~\cite{Green:2004wb, Carr:2016drx}.

Once formed, the PBHs can slowly decay by emitting Hawking radiation. In particular, PBHs with $M_{PBH} \lesssim 10^{11}~\text{kg}$ decay into thermal radiation before the onset of big bang nucleosynthesis and their abundance can thus be very large~\cite{Anantua:2008am,Hooper:2019gtx}. On the other hand, PBHs with $10^{11}~\text{kg} \lesssim M_{PBH} \lesssim 10^{14}~\text{kg}$ have a life-time comparable with the age of the universe and their abundance is highly constrained by the non-observation of their Hawking radiation. Heavier black holes are stable and contribute to dark matter, their abundance is constrained by the observed dark matter abundance as well as by direct searches, see e.g.\ Refs.~\cite{Carr:2009jm,Inomata:2017okj} for an overview.

For a given amplitude of the scalar power spectrum, the probability of forming PBHs depends on the statistical properties of the scalar fluctuations, since typically PBH formation is a rare event occurring in the tail of the distribution function. For a gaussian distribution any power spectrum generating stable black holes with $\langle \zeta^2 \rangle \gtrsim 10^{-2}$ leads to an overclosure of the universe~\cite{Kalaja:2019uju}. For a positive $\chi^2$-distribution, as expected for the sourced scalar perturbations in axion inflation, this value is lowered to $\langle \zeta^2 \rangle \gtrsim  10^{-3}$~\cite{Linde:2012bt}.
The amplitude of the power spectrum in Fig.~\ref{fig:PSnum} clearly exceeds these values towards the end of inflation. Thus requiring $M_{PBH}(N) < 10^{11}~\text{kg}$ to avoid these overclosure bounds restricts the enhancement of the scalar power spectrum to the last $\sim 10$ e-folds, see Eq.~\eqref{eq:PBHmass}. Here we have set $j = 3$ since the expected large abundance of PBHs generated right after inflation will lead to an early matter dominated phase.

Consequently, the power spectrum depicted in Fig.~\ref{fig:PSnum} which is only enhanced in the last $\sim 5$ (9) e-folds for $f = 1/20$ (1/25), is (marginally) compatible with bounds from PBH formation. Significantly larger values of $1/f$ will lead to overproduction of stable PBHs, though the precise bound will depend on the details of the last stages of inflation, see discussion below Eq.~\eqref{eq:powerspectrum_et}.
On the contrary, a large abundance of metastable black holes as found for $1/f \lesssim 25$ entails several interesting phenomenological consequences. Firstly, an early PBH dominated phase, eventually releasing its energy into thermal Hawking radiation, provides a remarkable reheating mechanism. Any radiation released during preheating or in the inflaton decay is strongly red-shifted during the PBH dominated era, and hot big cosmology is re-ignited once the PBHs decay. Among others, this poses interesting challenges for baryogenesis. 
Secondly, there are three significant sources of gravitational waves (GWs): (i) GWs sourced by the gauge field population during inflation~\cite{Cook:2011hg}, (ii) GWs sourced (at second order) from the large scalar perturbations~\cite{Mollerach:2003nq,Ananda:2006af,Baumann:2007zm} and (iii) GWs sourced as a component of the Hawking radiation of the decaying PBHs~\cite{Anantua:2008am,Dong:2015yjs}. All of these sources result in high frequency ($\sim$ MHz and beyond) GWs, beyond the scope of current experiments but suggesting a potential target for potential future high frequency experiments. We expect that the characteristic oscillating features of the source $\langle \vec E \vec B \rangle$ will also be visible in the GW spectrum. Note that any GWs which are sub-horizon during the PBH dominated phase will be strongly diluted, leading to an interesting interplay between the GW and PBH spectrum. This applies in particular to GWs generated during preheating right after inflation~\cite{Adshead:2019lbr}.


\section{Conclusions \label{sec:conclusions}}

Axion inflation is generically accompanied by an explosive gauge field production, triggered by a tachyonic instability of roughly horizon sized gauge field modes, which is in turn sourced by the inflaton velocity. The energy budget of this gauge field configuration is drained from the kinetic motion of the inflation, which can be described as a backreaction of the classical gauge fields on the homogeneous inflaton equation of motion. In this paper we study the resulting coupled system of differential equations numerically, pointing out several new aspects which point to a more complex dynamics than previously anticipated.

The tachyonic instability is most effective on slightly sub-horizon scales, and hence the amplitude of any gauge field mode is set by the value of the inflaton velocity just before this mode crosses the horizon. On the other hand, the non-linear backreaction term is dominated by super-horizon gauge field modes, and hence reacts with a time lag to any change in the inflaton velocity. As the average speed of the inflaton increases over the course of inflation this system eventually hits a resonance frequency, where this time-lag corresponds to a phase shift of $\pi$. This leads to oscillations with increasing amplitude and fixed frequency in e-fold time, clearly visible in the inflaton velocity, the backreaction term and the gauge field energy density. This drastically changes the dynamics of axion inflation in the strong backreaction regime.

An example of an observable which is significantly impacted by this change in the inflaton dynamics is the scalar power spectrum. At very early times, when the scales relevant for the CMB exited the horizon, the backreaction is irrelevant and the spectrum closely resembles the usual spectrum of vacuum fluctuations. On smaller scales, corresponding to later stages of inflation, the scalar power spectrum receives an additional contribution sourced by the inhomogeneous part of the gauge field distribution, leading to an enhancement by many orders of magnitude. In this paper we re-visit the equation of motion for the scalar perturbations, reproducing results found previously in the weak backreaction regime but finding a significant larger amplitude for the scalar power spectrum in the strong backreaction regime. This result holds even when working with a time-averaged backreaction, i.e.\ discarding the resonance discussed above. Including the resonance leads to additional oscillatory features in the power spectrum at small scales. However, our results also indicate that the strong backreaction regime entails such large scalar perturbations (invoking in particular significant spatial gradients in the inflaton field) that the perturbative description fails. The formation of (metastable) primordial black holes seems unavoidable, entailing interesting phenomenological consequences, but any more quantitative analysis requires a non-perturbative description which is beyond the scope of the present paper.

In this context, it is interesting to note the recent progress made in simulating the preheating phase of this model on the lattice~\cite{Adshead:2015pva,Cuissa:2018oiw,Adshead:2019lbr} (see also \cite{Agrawal:2018vin} for related work). The challenges induced by the growing separation of scales in an expanding Universe limits the amount of e-folds which can be tracked, but the characteristic time scale $\Delta N_\xi \simeq \ln(\xi^2/2)$  of the resonance seems to be within reach of such analyses.  The preheating phase, and in particular its gravitational wave production, can impose stringent bounds on the axion to photon coupling, down to $1/f \lesssim 10$~\cite{Adshead:2019lbr}. However, an early PBH dominated phase, triggered by the drastically enhanced scalar power spectrum, would significantly dilute the energy density in gravitational wave radiation which redshifts faster than the PBH component. This could re-open the parameter space of larger couplings. We leave a more detailed study of this question to future work.

The observed resonance phenomenon will not only affect the scalar power spectrum but also the tensor power spectrum, since it too receives a contribution sourced by the gauge field population. Moreover, we expect that similar resonance phenomena can occur in other cosmological systems which feature a tachyonic instability of gauge fields modes driven by a non-vanishing axion velocity. This includes models of baryogenesis driven by the motion of axion-like particle~\cite{Jimenez:2017cdr,Domcke:2019mnd} and models of cosmological relaxation of the electroweak scale utilizing gauge field friction~\cite{Hook:2016mqo,Choi:2016kke,Tangarife:2017vnd,Tangarife:2017rgl,Fonseca:2018xzp,Fonseca:2019lmc}. We leave these questions to future work.

\paragraph{Acknowledgements}
It is a pleasure to thank Daniel Figueroa, Ryo Namba and Evangelos Sfakianakis for helpful discussions related to this project. Special thanks to Lorenzo Sorbo for providing valuable comments on the manuscript.
This work was partially funded by the Deutsche Forschungsgemeinschaft under Germany's Excellence Strategy - EXC 2121 ``Quantum Universe'' - 390833306. AW and YW are supported by the ERC Consolidator Grant STRINGFLATION under the HORIZON 2020 grant agreement no. 647995.


\appendix

\newpage

\section{Phase shift}\label{app:phaseshift}
In this appendix we derive in a slightly different manner the value of the characteristic time scale  $\Delta N_\xi$ that denotes the lag between $\langle \vec E \vec B \rangle (N) $ and $\xi(N)$, given in Eq. \eqref{eq:LagEstimate}. \\

\begin{figure}[t!]
\includegraphics[width=0.8\textwidth]{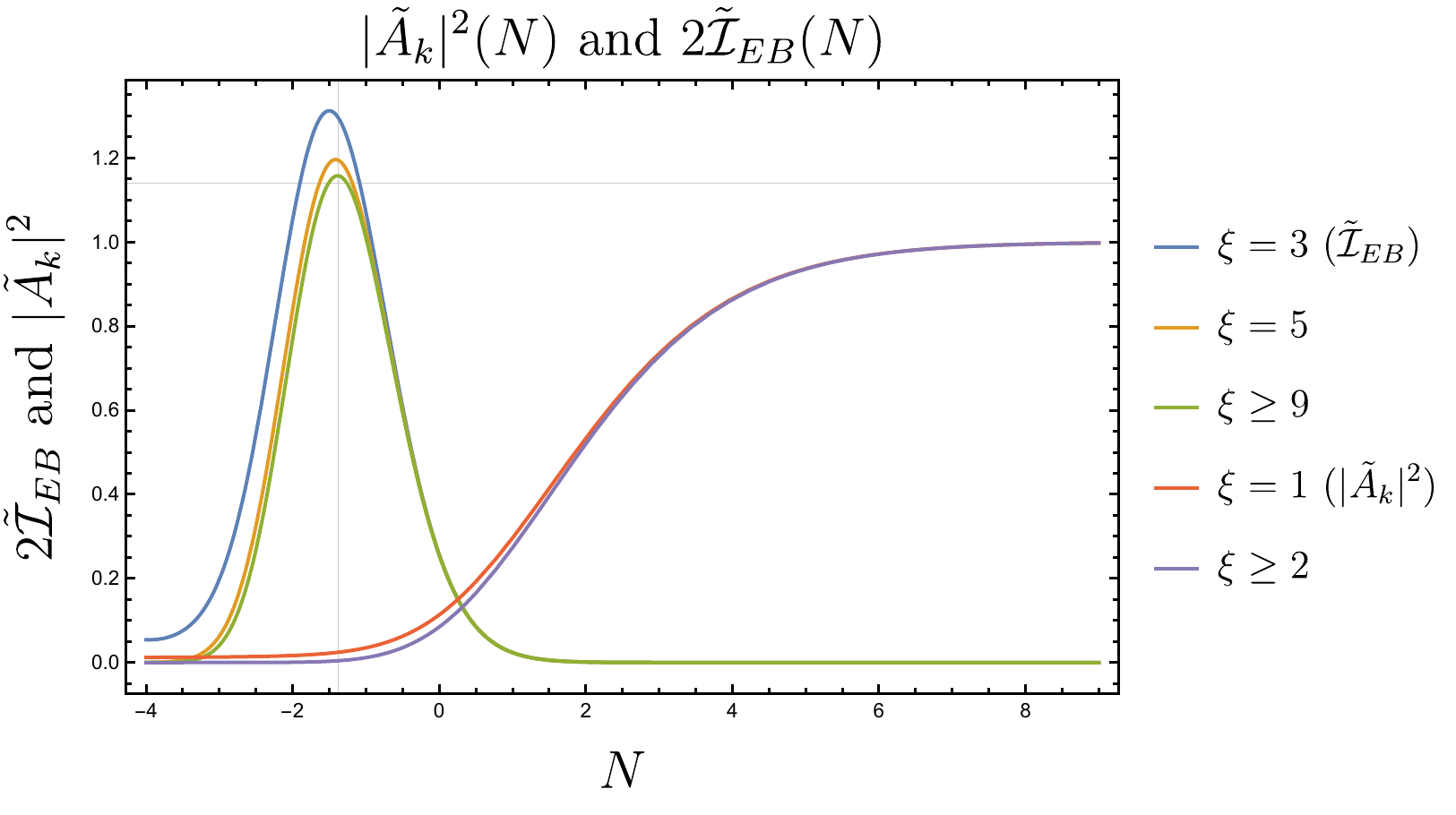}
\centering
\caption{Evolution of $\left|\tilde A_k\right|^2 (N)$ evaluated at $N_k =0$ (red and purple lines) and support of $2\tilde{\mathcal{I}}_{E B}(N)$ (blue, orange and green lines) for various values of constant $\xi$. }
\label{fig:selfsimilar}
\end{figure}

First, we notice that in the case of constant $\xi$ we can define a self-similar function $\tilde A(N)$ that captures the growth of the gauge modes for any large enough value of $\xi$. If we evaluate the enhanced gauge modes $A_{-\lambda}(N, k)$ at the time $N+\ln 2\xi$ and additionally rescale their amplitude with $\sqrt{\frac{2 \pi k \xi}{e^{\pi\xi} \sinh(\pi \xi)}}$ (such that they asymptote to unity) their equation of motion in e-folds reads
\begin{equation}
\tilde A_k^{\prime\prime} + \tilde A_k^\prime + \frac{k}{ aH}\left(\frac{k}{4 aH \xi^2} - 1\right)\tilde A_k = 0\ .
\label{eqn:selfsimilareomA}
\end{equation}
Therefore, plugging in the constant $\xi$ solution for the gauge modes given in Eq. \eqref{eq::AkWhit}, we find that
\begin{equation}
\tilde A_k = \sqrt{\frac{\pi \xi}{\sinh(\pi \xi)}} W_{-i\xi, 1/2}\left(-i\frac{k}{a H \xi}\right)
\end{equation}
is a `self-similar' solution that only depends on $N$ (and on a trivial way on $k$) as long as the $k/4aH\xi^2$ correction can be neglected in Eq. \eqref{eqn:selfsimilareomA}.
Numerically, we find indeed that the $\xi$-dependence drops out for $\xi \gtrsim 2$.  See Figure \ref{fig:selfsimilar}. The original gauge mode $A_k$ can then be expressed in terms of $\tilde A_k$ as
\begin{equation}
A_k(\xi,N) = \sqrt{\frac{\sinh(\pi\xi)}{2\pi k \xi}}e^{\pi \xi/2} \tilde A_k(N-\ln 2\xi)\ .
\end{equation}
Similarly, using Eq. \eqref{eq:EB}, we define a self-similar function for the integrand of $\langle \vec{E}\vec{B} \rangle$
\begin{equation}
\tilde{\mathcal{I}}_{E\cdot B}(N) = \frac{\pi \xi e^{-3N}}{\sinh(\pi\xi)}\partial_N |W_{-i\xi, 1/2}\left(-2i e^{-N-\ln 2\xi} \right)|^2,
\label{eqn:selfsimilarI}
\end{equation}
such that the integrand of $\langle \vec{E}\vec{B} \rangle$ (in $d \ln k$) is given by
\begin{equation}
{\mathcal{I}}_{E\cdot B}(k,\xi, N) =\frac{H^4 \sinh(\pi\xi) e^{\pi\xi}}{64 \pi^3 \xi^4 a_0^3}\tilde{\mathcal{I}}_{E\cdot B}(N - \ln 2\xi -N_k) \ ,
\end{equation}
where $N_k$ is the time that the mode $k$ crosses the horizon.
The self-similar integrand \eqref{eqn:selfsimilarI} indeed becomes independent of $\xi$, but only for  $\xi \gtrsim 4$. This is because of the additional $e^{-3N}$ that shifts the peak almost 3 e-foldings to sub-horizon scales. We find that $\tilde{\mathcal{I}}_{E\cdot B}$ peaks at $ N \approx -1.38 \approx \ln (1/4)$ with amplitude $\tilde{\mathcal{I}} \approx 0.57$ and has most of its support $ \pm 1.5$ e-foldings around it. See Figure \ref{fig:selfsimilar}.
The integrand therefore peaks approximately at the wavenumber that crosses the horizon at $N_\text{peak} = N - \ln\xi/2$.  \\

Second, when $\xi$ is time-dependent, the gauge mode function $A_k$ grows to a 
plateau value, corresponding to the value reached for some constant $\xi_\text{eff}$, $A_k(\xi(\tau); k \tau \ll 1) = A_k(\xi_\text{eff}; k \tau \ll 1)$. If $\xi$ is slowly varying in time, we expect $\xi_\text{eff}$ to track $\xi$ adiabatically with some time delay. Indeed, we find that a good fit is given by 
\begin{equation}
\xi_\text{eff}(N_k) = \xi(N_\ast) \quad \text{with} \quad N_\ast = N_k - \log(\xi(N_\ast)/a),
\label{eqn:xieff}
\end{equation}
where $N_\ast$ is implicitly defined and $a\approx 1.2-2.0$. This refines the argument given in Sec.~\ref{sec:toymodel} that the value of $\xi$ at $k/aH\simeq\xi$ determines the growth of $A_k$. 
If we deviate from adiabatic tracking, however, the effective $\xi$ averages out to some degree. 
This makes sense, as the growth of the gauge modes will start to feel a range of values of $\xi$. As we can see from Fig.~\ref{fig:StepFunctionLag}, the effective $\xi$ that $\langle \vec E \vec B \rangle$ feels is not exactly the value of $\xi$ evaluated at a particular instance of time, but rather an average over a range of values. We can imagine a smoothing window of width $\sim \ln 4\xi^2$ going over the dashed curve as time proceeds. Only if the smoothing window has completely passed the jump at $N_0$, then $\langle \vec E \vec B \rangle$ will have reached its final plateau value. Therefore, we expect that Eq. \eqref{eqn:xieff} needs to be refined if $\xi$ changes considerably over the coarse of $\sim \ln 4\xi^2$ e-folds.\\

\begin{figure}[t!]
\includegraphics[width=0.8\textwidth]{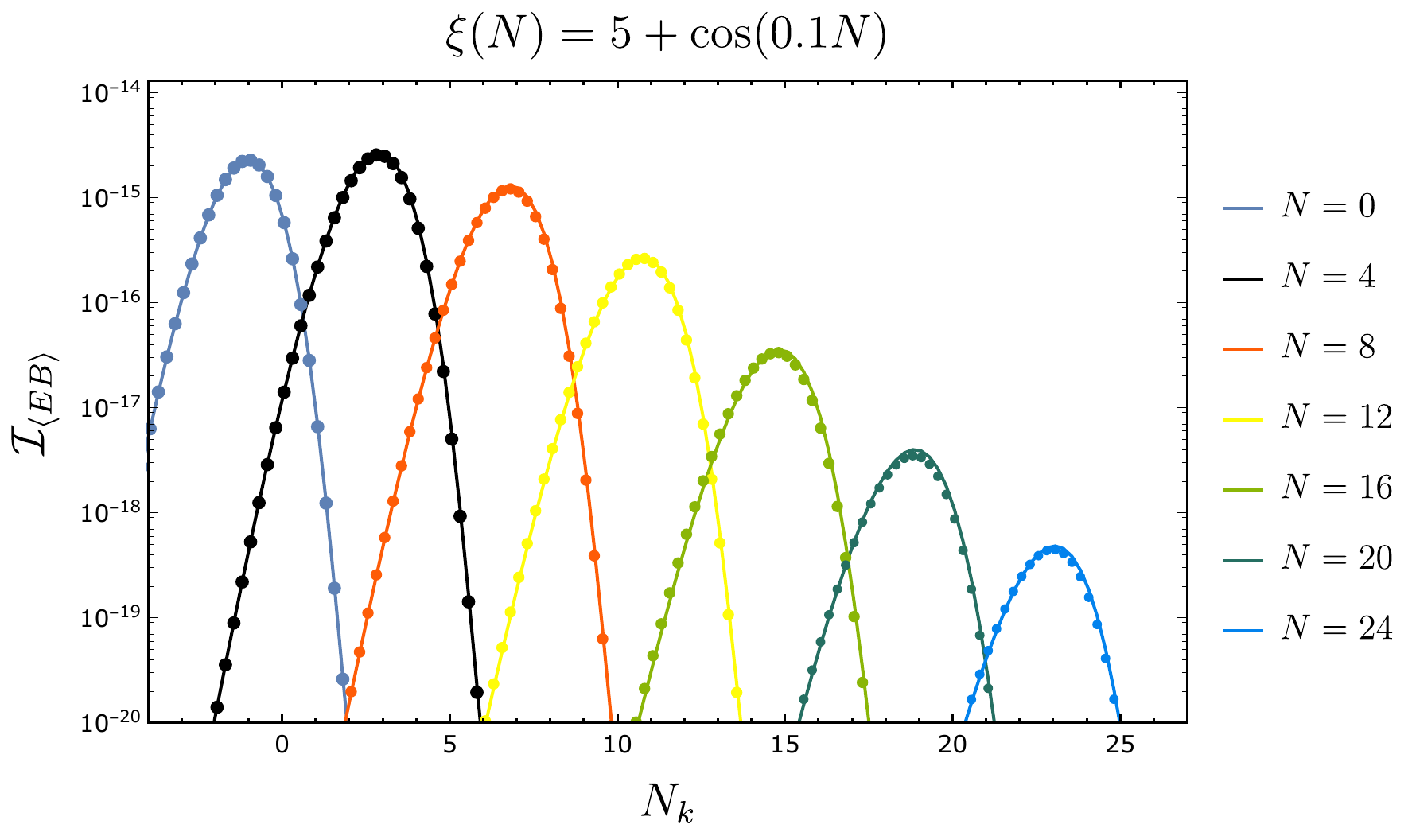}
\centering
\caption{Comparison of the integrand of $\langle \vec{E}\vec{ B}\rangle$ (discrete points) with $\mathcal{I}_{E B}(N_k, \xi_\text{eff}(N_k), N)$ (solid lines) for an oscillating $\xi = 5 + \cos (0.1 N) $ evaluated at various times $N$. }
\label{fig:integrandEBxieff}
\end{figure}

At this point we make an ansatz: the integrand of $\langle \vec{E}\vec{B} \rangle$ is given by $\mathcal{I}_{E\cdot B}(k,\xi_\text{eff}(N_k), N)$. This indeed seems to be a good approximation for slowly varying $\xi$, see Figure \ref{fig:integrandEBxieff}, where we take $a=1.45$.
The above considerations allow us to find a semi-analytical estimate for $\Delta N_\xi$. Let us focus on the harmonic
\begin{equation}
\xi(N) = \bar\xi + A \cos (\omega_\xi N)\ .
\end{equation}
The first maximum of $\xi_\text{eff}$ reflecting the maximum of $\xi$ at $N_\ast=0$ will be at 
\begin{equation}
0 = N_\text{max} - \ln\left((\bar\xi + A)/a\right) \quad \longrightarrow \quad N_\text{max} = \ln\left((\bar\xi + A)/a\right)\ .
\end{equation}
Meanwhile, the integrand of $\langle \vec{E}\vec{B} \rangle(N)$ peaks at 
$N_p = N - \ln(\xi_\text{eff}(N_p)/2)$ and will take the maximal value at $N=\Delta N_\xi$ when $N_\text{peak} = N_\text{max}$, hence 
\begin{equation}
N_\text{max} = \Delta N_\xi - \ln\left((\bar\xi + A)/2\right) \quad \longrightarrow \quad \Delta N_\xi   = \ln\left((\bar\xi + A)^2/2a\right)\ .
\end{equation}
We find that a good fit is given for $a \approx 1.45 $ and is shown in Figure \ref{fig:DeltaNvsXi} together with the original estimate $\Delta N_\xi = \ln(\xi^2/2)$ that was argued for in the main text.

\section{Details on the numerics \label{app:numerics}}
In order to obtain our numerical results we use an iterative procedure whose starting point is given by the analytical estimate of the mode function $A_k$ assuming constant inflaton speed $\phi'(N)$, Eq. (\ref{eq::AkWhit}):

\begin{equation}
\begin{array}{lll}
\langle \vec{E}\vec{B} \rangle_{(0)} = \frac{1}{2^{21}\,\pi^2}  \frac{H_0^4}{\xi^4} e^{2\pi\xi}\int_0^{8\xi}x^7 e^{-x}dx \,,\\[10pt]
\end{array}
\label{eq:EB0}
\end{equation}
\begin{equation}
\begin{array}{lll}
\langle\rho_{_{EB}}\rangle_{(0)} =\langle\frac{ E^2+B^2}{2} \rangle_{(0)} =\frac{6!}{2^{19}\pi^2}\frac{H_0^4}{\xi^3}e^{2\pi\xi}\,,
\label{eq:EEBB0}
\end{array}
\end{equation}
where $H_0$ is given by the Hubble parameter in absence of any backreaction, $H_0^2=\frac{V(\phi)}{3-\frac{1}{2}\phi'^2}$.
Denoting the $j$-th order iteration quantities with the subscript $j$, our first step is to find the solution of the following differential equation for a given $\langle \vec E \vec B \rangle_{(j-1)}$ obtained in the previous iteration:
\begin{equation}
\phi_{(j)}''+(3-\epsilon_{(j)})\phi_{(j)}' +\frac{1}{H_{(j)}^2}\left( V_\phi(\phi_{(j)})+\frac{\alpha}{\Lambda}\langle\vec{E}\vec{B} \rangle_{(j-1)}\right)=0 \,,
\end{equation}
where
\begin{equation}
H_{(j)}^2=\frac{V(\phi_{(j)}) +\langle\rho_{_{EB}}\rangle_{(j-1)}}{3-\frac{\phi_{(j)}'^2}{2}}\,;\qquad\qquad \epsilon_{(j)}=\frac{1}{2}\phi_{(j)}'^2 +\frac{2}{3 H_{(j)}^2} \langle\rho_{_{EB}}\rangle_{(j-1)}\, .
\end{equation}
Once we get the solution of this equation, $\phi_{(j)}(N)$, we plug the derived quantities $H_{(j)}(N)$, $\epsilon_{(j)}(N)$ and $\xi_{(j)}(N)$ inside the gauge mode equations
\begin{equation}
\begin{array}{lll}
A_{k,\,\pm}''+ (1-\epsilon_{(j)})A_{k,\,\pm}'+\frac{k}{a H_{(j)}}\left(\frac{k}{a H_{(j)}}\mp 2\, \xi_{(j)}(N)\right)A_{k,\,\pm}=0 \,.
\end{array}
\end{equation}
Then, choosing an array of $k$-modes with an exponential spacing,  we  estimate the discretized version of $\langle\rho_{_{EB}}\rangle_{(j)}$ and $\langle \vec{E}\vec{B} \rangle_{(j)}$
\begin{equation}
\langle\rho_{_{EB}}\rangle_{(j)}= \frac{1}{4\pi^2a^4}\displaystyle\sum_{i=1}^{M} d\ln k_i \,\left( k_i^3  a^2 H_{(j)}^2 |A_{k_i}^{'\sigma}|^2+ k_i^5 |A_{k_i}^{\sigma}|^2-k_i^4\right)\theta\left(N-N_i\right) \,,\\[10pt]
\label{eq:rhoeebb}
\end{equation}
\begin{equation}
\langle \vec{E}\vec{B} \rangle_{(j)}=\sigma \frac{H_{(j)}}{4\pi^2 a^3}\displaystyle\sum_{i=1}^{M} d\ln k_i\,k_i^3\frac{\partial}{\partial N}|A_{k_i}^{\sigma}|^2 \theta\left(N-N_i\right) \,,\\[10pt]
\label{eq:rhoeb}
\end{equation}
where $\sigma$ is the polarization which experiences the tachyonic behaviour and the third term in Eq. (\ref{eq:rhoeebb}) accounts for the subtraction of the Bunch-Davies contributions. With $N_i=\min_N\left\{2 a H \xi -k_i <0 \right\}$ the  Heaviside $\theta$ function is introduced to take into account only those modes that have already become tachyonic.

The array of k-modes is defined as $k_p=k_{in}e^{\sum_{i=1}^{p-1}\Delta_i}$ where $p=2\dots M$, $k_1=k_{in}$ is the lowest momentum taken into account and $\Delta_i=\{0.1,0.02\}$. The value we choose for $\Delta_i$ depends on the oscillatory behaviour of the solution: the stronger the backreaction, the thinner the momentum grid. Given this choice, we can write down the integration step as $dk=k\,  d\ln k $. The weight related to the contribution of a single mode to the integral is evaluated using the trapezoidal rule, i.e. $ d\ln k_p=\frac{1}{2} \log\left(\frac{k_{p+1}}{k_{p-1}}\right)=\Delta_{p}$ and  $ d\ln k_1=\frac{1}{2} \log\left(\frac{k_{2}}{k_{1}}\right)=\frac{\Delta_1}{2}$,  $ d\ln k_M=\frac{1}{2} \log\left(\frac{k_{M}}{k_{M-1}}\right)=\frac{\Delta_{M-1}}{2}$. 

Once we have evaluated the integrals~\eqref{eq:rhoeebb} and \eqref{eq:rhoeb} in this way we are able to define next iteration quantities $\epsilon_{(j+1)}$,  $H_{(j+1)}$ and the new approximated equation of motion that the inflaton field needs to satisfy. Iterating this procedure allows us to find better approximations of the real solution of the system.  We stop the calculations when there is no appreciable difference between the consecutive iterations. We do not prove here that this procedure always converges at a reasonable rate. But if convergence is reached (as is the case in our explicit numerical examples), this procedure ensures a self-consistent solution of the integro-differential system (\ref{eq:eom_phi_0}), (\ref{eq:eom_A}) and (\ref{eq:EB}).

During the algorithm we check that the contributions coming from the non-tachyonic polarizations is completely negligible.

\section{Estimate of non-equal time correlation function \label{app:NETcorr}}

\subsection{Analytical estimate}
\label{app:app:NETcorr-a}

Far away from the resonance region the parameter $\xi$ varies only slowly and we can estimate the importance of the non-equal time contributions to Eq.~\eqref{eq:PS_masterformula} by looking at the result Eqs.~(A3) and (A4) from \cite{Anber:2009ua}. These expressions are based on parametrizing the gauge field mode functions with Whittaker functions, see Eq.~\eqref{eq::AkWhit}. Massaging the expressions a little bit, we get for the correlator
\begin{align}
\int d^3x \, e^{i\vec p \vec x} \, \langle0|\delta_{EB}(N',0)& \delta_{EB}(N'',x)|0\rangle \sim \frac{H'H''}{a'^3a''^3}\frac{e^{2\pi(\xi'+\xi'')}}{(\sqrt{\rho'}+\sqrt{\rho''})^{10}} C(\kappa)\quad,\quad\\
\nonumber\\
C(\kappa)=&\kappa^5\int\limits_0^\infty dq q^3 \int\limits_{-1}^1 d\alpha\sqrt{1+q^2+2q\alpha}e^{-\sqrt\kappa (q+\sqrt{1+q^2+2q\alpha})}\nonumber\\
&\times \left(1+\frac{\sqrt q}{(1+q^2+2q\alpha)^{1/4}}\right)\int\limits_0^{2\pi}d\phi|\epsilon_+(-\hat{q})\cdot \epsilon_+(\hat{q}+\hat{e}_z)|^2\quad.\nonumber
\end{align}
Here, we define $\rho\equiv 2\xi/(aH)$, $\vec q \equiv \vec k/|p|$, $\vec p \equiv |p|\hat e_z$ and $\kappa \equiv 4|p|(\sqrt{\rho'}+\sqrt{\rho''})^2$. {In this appendix only, for notational brevity} the superscripts $()'$ and $()''$ denote the given quantity at time $N'$ and $N''$, respectively.
We now see, firstly, that the $\kappa^5$ factor inside $C(\kappa)$ and the factor $1/(\sqrt{\rho'}+\sqrt{\rho''})^{10}$ multiplying $C(\kappa)$ cancel each other. Secondly, we recognize that the correlator is bounded from above by its value on far super-horizon scales $\kappa\to 0$, and that the correlator depends only polynomially on $a'$ and $a''$ in this limit. Hence we find that the correlator at late times scales as 
\begin{equation}
\int d^3x \, e^{i\vec p \vec x} \, \langle0|\delta_{EB}(N',0)\delta_{EB}(N'',x)|0\rangle \sim \frac{1}{a'^3a''^3}\sim e^{-3(N'+N'')}=e^{-6N'}e^{-3\Delta N} \,,
\end{equation}
assuming $N''>N'$ with loss of generality. By comparison, we conclude that the argument of $g(N, \Delta N)$ scales as $e^{-3\Delta N}$. 
For a functional form $f(\Delta N)=\exp(-c \, \Delta N)$  the integral
\begin{align}
g(N', 0) = \int_{N'}^\infty dN'' \exp(-c \, \Delta N) =  \frac{1}{c} 
\end{align}
is of ${\cal O}(1)$ for ${\cal O}(1)$ values of $c$. Since in our case we have $c=3$,  the inclusion of unequal time correlations does not significantly alter our result. This can also be confirmed by a comparison of our results with previous analysis~\cite{Anber:2009ua,Barnaby:2011vw} which included this unequal time correlator, see App.~\ref{app:comparison}.

\subsection{Numerical evaluation}

In the resonant regime the Whittaker functions used in App.~\ref{app:app:NETcorr-a} are no longer a good approximation to the full mode functions. In this region, we evaluate the non-equal time correlator numerically, based on the mode functions obtained in Sec.~\ref{sec:numerics}. 

In order to compute the shape of the non-equal time correlation function, we define symmetrized version of $\delta_{EB}$ (see Eq.~(\ref{eq:deltaphi_1})), analogous to the symmetrized $\langle \vec E \vec B \rangle$ introduced in  Eq.~(\ref{eq:EB}) (see also~\cite{Jimenez:2017cdr}),
\begin{align}
\delta_{EB}(\tau,x)_S&=\left(E^i(\tau,x) B^i(\tau,x)\right)_S - \langle \vec{E}\vec{B}\rangle_S(\tau') \nonumber \\
&=\frac{1}{2}\left( E^i(\tau,x) B^i(\tau,x)+ B^i(\tau,x) E^i(\tau,x) \right)- \langle \vec{E}\vec{B}\rangle_S(\tau) \,,
\end{align}
and consequently 
\begin{align}
\langle 0|\left[\delta_{EB}(\tau',x)_S \right. & \left. \delta_{EB}(\tau'',0)_S \right]_S|0\rangle \nonumber \\
& = \frac{1}{2}\langle 0| \delta_{EB}(\tau',x)_S\delta_{EB}(\tau'',0)_S+\delta_{EB}(\tau'',0)_S \delta_{EB}(\tau',x)_S\rangle|0\rangle \,.
\end{align}
If we consider only positive helicity modes, $\lambda=+$, and we use the following short notation
\begin{align}
E_1^i &=E^i(k,\tau'',\vec{x},+)\,, \quad 
B_1^i=B^i(k,\tau'',\vec{x},+)\,, \quad
E_2^j=E^j(k,\tau',0,+)\,, \quad \nonumber \\
B_2^j &=B^j(k,\tau',0,+)\,, \quad 
B_2^j=B^j(k,\tau',0,+)\,, \quad
A_+(\tau,\vec{k})=A(\tau,\vec{k})
\end{align}
we end up with

\begin{align}
&\displaystyle \int d^3\vec{x} e^{i \vec{q}\cdot \vec{x}}\langle 0|\left[\delta_{EB}(\tau',x)_S\delta_{EB}(\tau'',0)_S \right]_S|0\rangle=\nonumber\\[3pt]
&\quad=\displaystyle \frac{1}{2}\int d^3\vec{x} e^{i \vec{q}\cdot \vec{x}}\left[\langle E_1^iE_2^j\rangle \langle B_1^iB_2^j\rangle+ \langle E_1^iB_2^j\rangle \langle B_1^iE_2^j\rangle+\right.\nonumber\\
&\qquad\qquad\qquad\qquad\qquad\qquad\left.+\langle E_2^jE_1^i\rangle \langle B_2^j B_1^i\rangle+\langle E_2^jB_1^i\rangle \langle B_2^jE_1^i\rangle \right]\nonumber\\[3pt]
&\quad=\displaystyle \frac{1}{2a'^4a''^4}\int \frac{d^3\vec{k}}{(2\pi)^3}|\vec{k}|^2 \left| \vec{\epsilon}_+(\vec{k})\cdot \vec{\epsilon}_+(-\vec{k}-\vec{q})\right|^2\times \nonumber\\
&\qquad\times\Bigg\{ \partial_\tau A(\tau',-\vec{k}-\vec{q})\partial_\tau A^{*}(\tau'',-\vec{k}-\vec{q})A(\tau',\vec{k})A^{*}(\tau'',\vec{k})\,+\nonumber\\
&\displaystyle \qquad\quad +2 \frac{|-\vec{k}-\vec{q}|}{|\vec{k}|} Re\left[ \partial_\tau A(\tau',-\vec{k}-\vec{q})A^{*}(\tau'',-\vec{k}-\vec{q})A(\tau',\vec{k})\partial_\tau A^{*}(\tau'',\vec{k}) \right]+\nonumber\\
&\displaystyle \qquad\quad \left. + \frac{|-\vec{k}-\vec{q}|^2}{|\vec{k}|^2}A(\tau'',-\vec{k}-\vec{q})A^{*}(\tau',-\vec{k}-\vec{q})\partial_\tau A(\tau'',\vec{k})\partial_\tau A^{*}(\tau',\vec{k})\right\} 
\label{Eq:FtransSS}
\end{align}

where in this appendix only, $a'\equiv a(\tau')$ and $a''\equiv a(\tau'')$.
Given that the positive polarization vector can written as
\begin{equation}
\epsilon_+(\vec{k})=\frac{\hat{k}\cdot \hat{e}_x +i\left(\hat{k}(\hat{k}\cdot \hat{e}_x)-\hat{e}_x\right)}{\sqrt{2} |\hat{k}\cdot \hat{e}_x|},
\end{equation}
if we assume that $\vec{q}=\{0,0,q\}$, we can see that using polar coordinates and setting $\cos(\theta)=\alpha$, the polarization dependent factor inside Eq.~(\ref{Eq:FtransSS}) becomes
\begin{align}
\left|\epsilon_+(\vec{k})\cdot\epsilon_+(-\vec{q}-\vec{k})\right|^2=&\frac{2k^2+4 k q \alpha+q^2(1+\alpha^2)}{4k^2\left(1+2\alpha\frac{q}{k}+\frac{q^2}{k^2}\right)}+\nonumber \\
&+ \frac{k^3 + 3k^2 q \alpha+q^3\alpha+kq^2(1+2\alpha^2)}{2k^3\left(1+2\alpha\frac{q}{k}+\frac{q^2}{k^2}\right)^{3/2}}.
\end{align}
In order to have a more compact notation we also define
\begin{align}
C_1(k,q,\alpha)=\left|\epsilon_+(\vec{k})\cdot\epsilon_+(-\vec{q}-\vec{k})\right|^2 \,,  \quad \nonumber \\
C_2(k,q,\alpha)= C_1(k,q,\alpha)\sqrt{1+2\alpha\frac{q}{k}+\frac{q^2}{k^2}} \,,
\end{align}
and since the gauge mode equation of motion depends just on the magnitude of the $k$-vector, we can write
\begin{align}
&A(N,\vec{k})=A(N,k);\nonumber \\
 &A(N,-\vec{k}-\vec{q})= A(N,k\sqrt{1+2\alpha\frac{q}{k}+\frac{q^2}{k^2}})\equiv A(N,k,q,\alpha).
\end{align}
Rearranging Eq.~\ref{Eq:FtransSS} and using the number e-foldings as time variable we get 
\begin{align}
\displaystyle \int &d^3\vec{x} e^{i \vec{q}\cdot \vec{x}}\langle 0|\left[\delta_{EB}(N',x)_S\delta_{EB}(N'',0)_S \right]_S|0\rangle=\displaystyle \frac{H'H''}{a'^3a''^3}\int_0^\infty \frac{dk}{(4\pi^2)}k^4\int_{-1}^{1}d\alpha \times\qquad\qquad \nonumber \\[5pt]
& \displaystyle  \left\{C_1(k,q,\alpha) Re\left[\partial_N A(N',k,q,\alpha)\partial_NA^{*}(N'',k,q,\alpha)A(N',k)A^{*}(N'',k)\right]\,\right.\nonumber\\[5pt]
&\displaystyle \left.\quad + C_2(k,q,\alpha) Re\left[ \partial_N A(N',k,q,\alpha)A^{*}(N'',k,q,\alpha)A(N',k)\partial_NA^{*}(N'',k) \right]\right\} 
\end{align}
where in this appendix only, $H'\equiv H(N')$ and $H'' \equiv H(N'')$. 
It is easy to see that the final result has the desired properties: it is real and symmetric under $N'\leftrightarrow N''$ and $\vec{x} \leftrightarrow -\vec{x}$.

As in App.~\ref{app:app:NETcorr-a} we focus on far super-horizon scales $q\to 0$,
\begin{align}
\langle \delta_{EB}(N')\delta_{EB}(N'')\rangle=&\displaystyle \lim_{q\rightarrow 0}\displaystyle \int d^3\vec{x} e^{i \vec{q}\cdot \vec{x}}\langle 0|\left[\delta_{EB}(N',x)_S\delta_{EB}(N'',0)_S \right]_S|0\rangle \nonumber \\
 =& \displaystyle \frac{H'H''}{a'^3a''^3}\int_0^\infty \frac{dk}{(2\pi^2)}k^4\times \nonumber \\
 &\quad \displaystyle  \left\{ Re\left[\partial_N A(N',k)\partial_NA^{*}(N'',k)A(N',k)A^{*}(N'',k)\right]\,+\right. \nonumber \\[3pt]
&\qquad \left. + Re\left[ \partial_N A(N',k)A^{*}(N'',k)A(N',k)\partial_NA^{*}(N'',k) \right]\right\} .
\end{align}
For numerical purposes we discretize the integral as follows
\begin{align}
\langle \delta_{EB}(N')&\delta_{EB}(N'')\rangle=\displaystyle \frac{H'H''}{(2\pi^2)a'^3a''^3} \sum_{k_i} d \ln k_i\, k_i^5 \sum_{j}\Delta\alpha \;\times \nonumber\\
&\left\{ Re\left[\partial_N A(N',k_i)\partial_N A^{*}(N'',k_i)A(N',k_i)A^{*}(N'',k_i)\right]\,+\right.\nonumber\\[3pt]
&\left.\displaystyle \quad + Re\left[ \partial_N A(N',k_i)A^{*}(N'',k_i)A(N',k_i)\partial_N A^{*}(N'',k_i) \right]\right\} .
\end{align} 
the discretization scheme is the same as in App. (\ref{app:numerics}).

\begin{figure}
\begin{center}
 \includegraphics[width=0.50\textwidth]{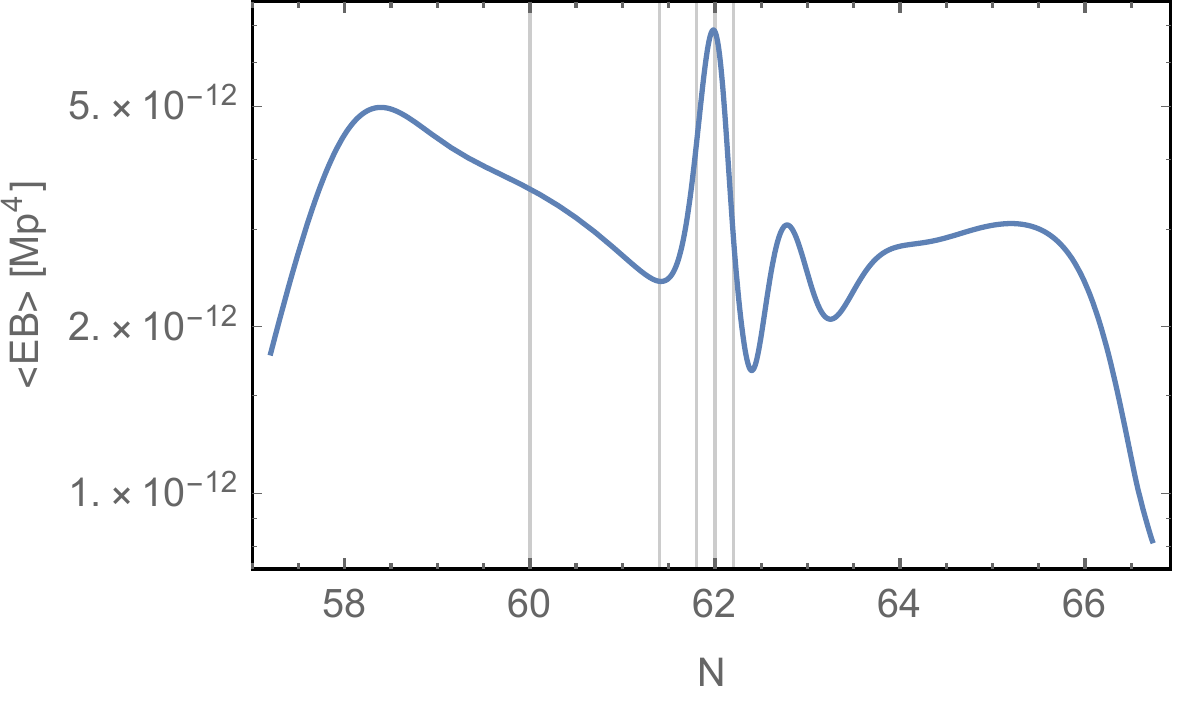} \hspace{2pt}\includegraphics[width=0.47\textwidth]{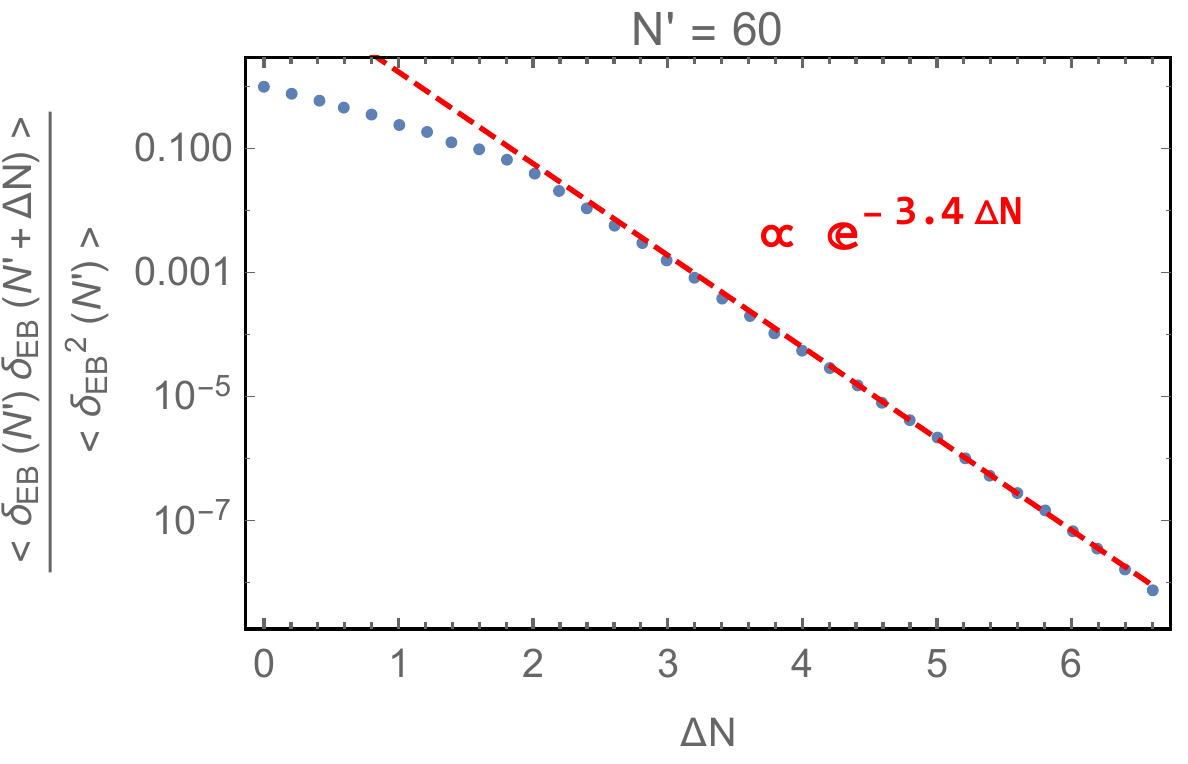}\\
\includegraphics[width=0.48\textwidth]{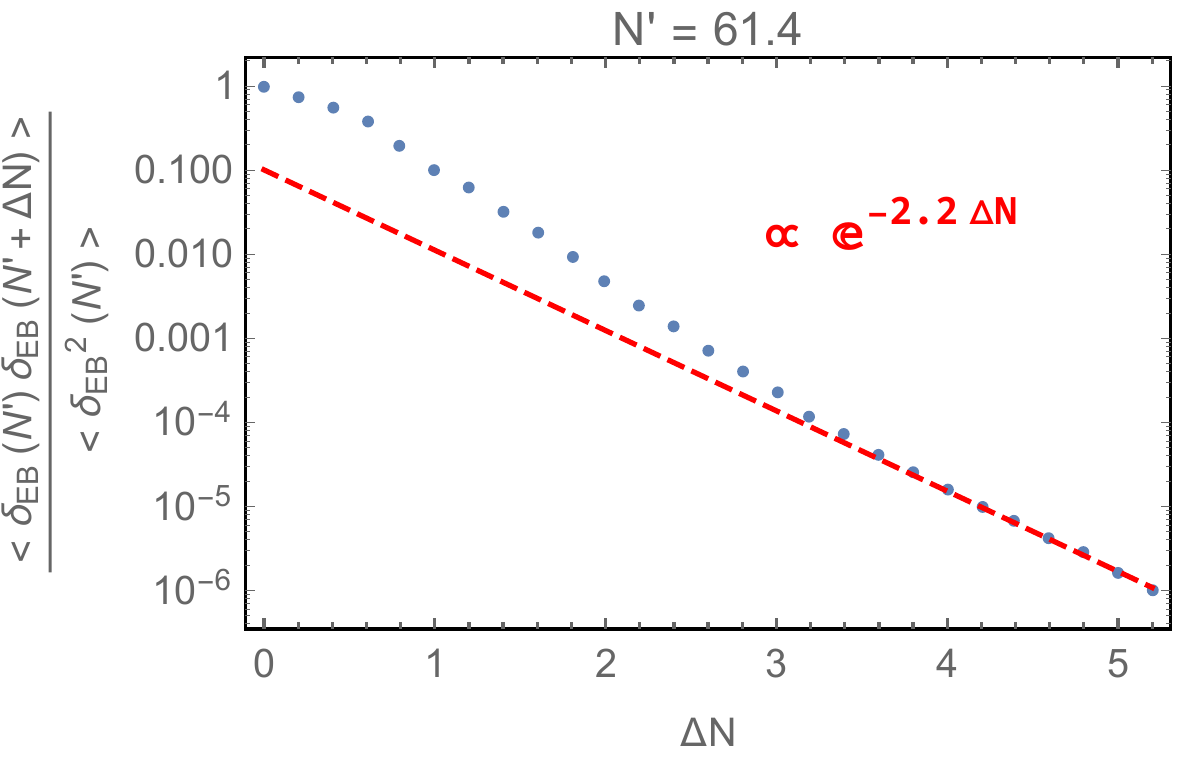}\hspace{5pt}\includegraphics[width=0.48\textwidth]{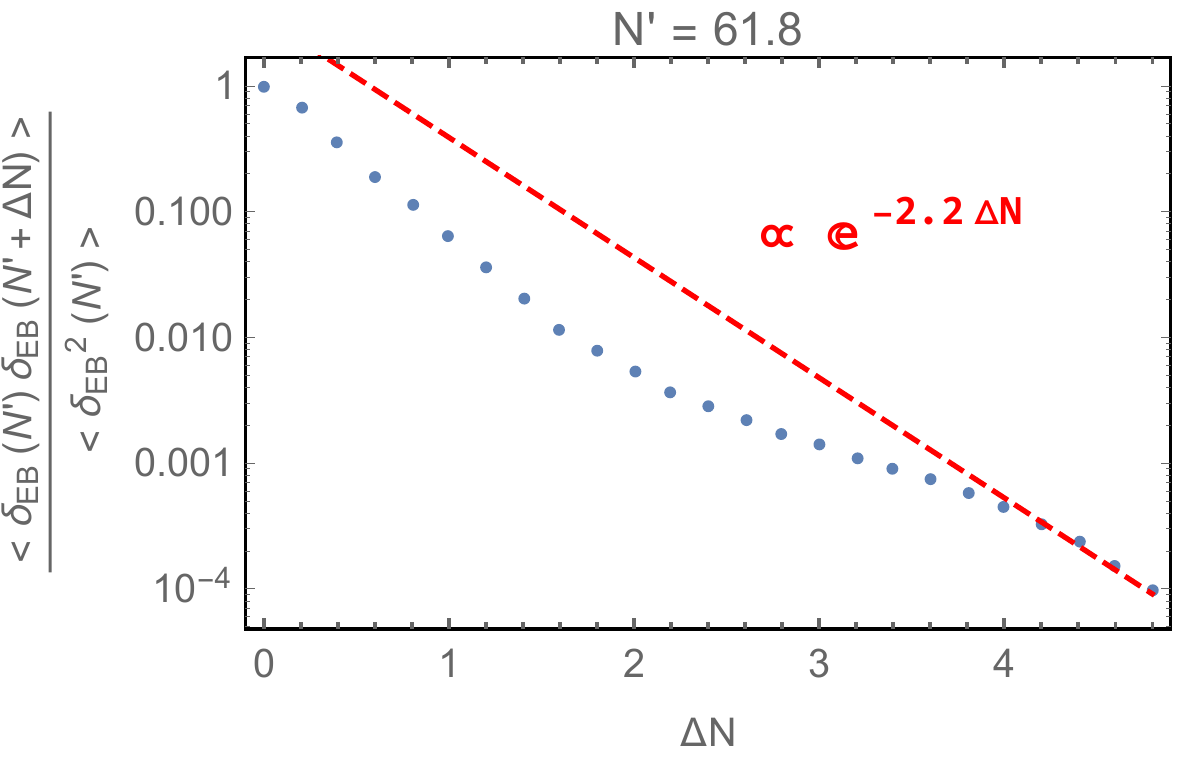}\\
\includegraphics[width=0.48\textwidth]{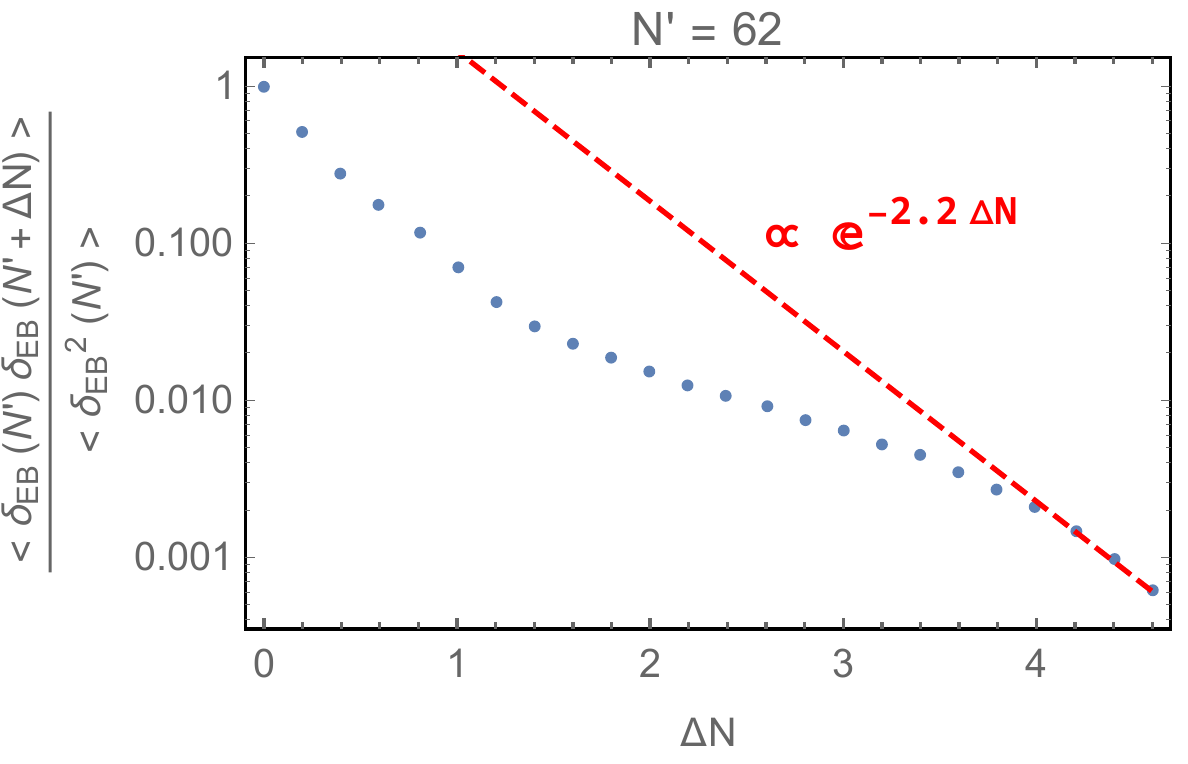}\hspace{5pt}\includegraphics[width=0.48\textwidth]{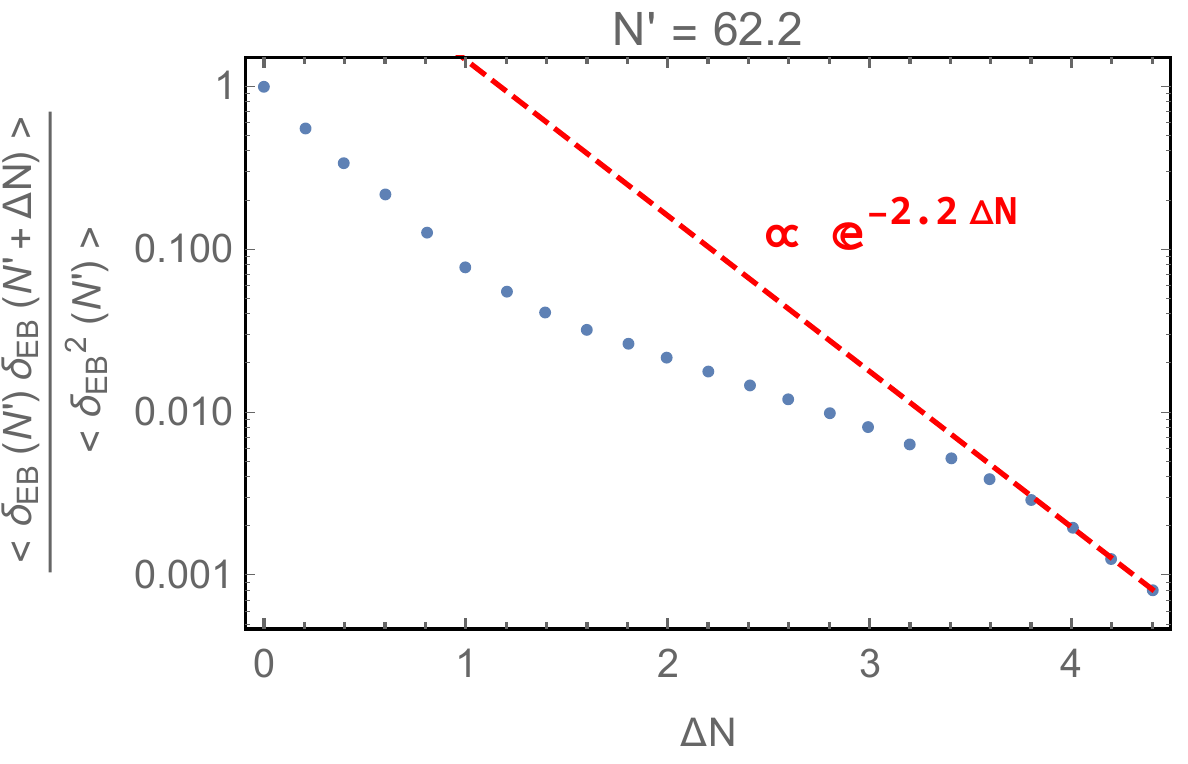}
\end{center}
\caption{Integrand of $g(N',\Delta N)$, for various values of $N'$ for $1/f = 25$. The top left panel corresponds to the bottom left panel of Fig.~\ref{fig:mesh1} and serves as an orientation to identify the position of the maxima and minima. All other panels show the integrand of $g(N', \Delta N)$ for local minima ($N'=61.4$), local maxima ($N'=62$) and steep regions ($N'=\{60,61.8,62.2\}$) of $\langle\vec{E}\vec{B}\rangle$. The red lines give the best fit for the exponentially decreasing tail of the distributions.}
\label{fig:NETcorr}
\end{figure}

We can now compute a numerical estimate of the normalized non-equal time correlation function that was introduced in Eq.~(\ref{eq:g_def}),
\begin{align}
 g(N', \Delta N) = \langle \delta_{EB}^2(N') \rangle^{-1} \int_{N' + \Delta N}^\infty dN'' \langle \delta_{EB}(N') \delta_{EB}(N'') \rangle \,.
\end{align}
Fig.~\ref{fig:NETcorr} shows the integrand of $g(N', \Delta N)$ at five distinct times deep in the resonance regime.
As in our analytical estimate in App.~\ref{app:app:NETcorr-a}, the integrand of $g(N', \Delta N)$ drops exponentially as $\exp(- c \Delta N)$ with $c = {\cal O}(1)$. For small values of $\Delta N$ the behaviour deviates from the exponential decay. Numerically performing the integral for some representative choices of $\Delta N$ yields
\begin{itemize}
\item{$N'=60$, $\Delta N=\{1,0.5,0.1\}$, $g(N',\Delta N)/\gamma= \{0.19,0.46,0.86\}$}
\item{$N'=61.4$, $\Delta N=\{1,0.5,0.1\}$, $g(N',\Delta N)/\gamma= \{6.6 \times 10^{-2},0.31,0.82\}$}
\item{$N'=61.8$, $\Delta N=\{1,0.5,0.1\}$, $g(N',\Delta N)/\gamma= \{6.4\times 10^{-2},0.24,0.76\}$}
\item{$N'=62$, $\Delta N=\{1,0.5,0.1\}$, $g(N',\Delta N)/\gamma= \{0.13,0.31,0.76\}$}
\item{$N'=62.2$, $\Delta N=\{1,0.5,0.1\}$, $g(N',\Delta N)/\gamma= \{0.15,0.34,0.79\}$}
\end{itemize}
We conclude that the support of $g(N', \Delta N)$ is mainly focused at small values of $\Delta N$, i.e.\ that unequal time correlations are mainly relevant on time scales over which $\langle \delta^2(EB) \rangle$ does not change too drastically. However, the contributions from more distant times are not fully negligible, and hence we expect ${\cal O}(1)$ corrections to the power spectrum in the resonance regime. These will tend to slightly smooth the maxima and minima of the power spectrum in this regime. However, since the minima are unobservable and the maxima violate perturbativity (see discussion in Sec.~\ref{sec:PS}) this does not significantly impact our discussion in the main text.

\section{Scalar power spectrum: comparison with earlier work \label{app:comparison}}

The scalar power spectrum generated during axion inflation has been previously estimated in Refs.~\cite{Anber:2009ua,Barnaby:2010vf,Barnaby:2011vw,Linde:2012bt} based on the analytical estimate for $\langle \vec E \vec B \rangle$ given in Eq.~\eqref{eq:EB_estimate}. In this appendix we briefly review these derivations and their limitations. Of particular interest to us are Refs.~\cite{Anber:2009ua,Barnaby:2011vw} which are based on the Greens function method. Generalizing this approach leads to the results for the power spectrum reported in the main text.

\begin{figure}
 \centering
 \includegraphics[width = 0.6 \textwidth]{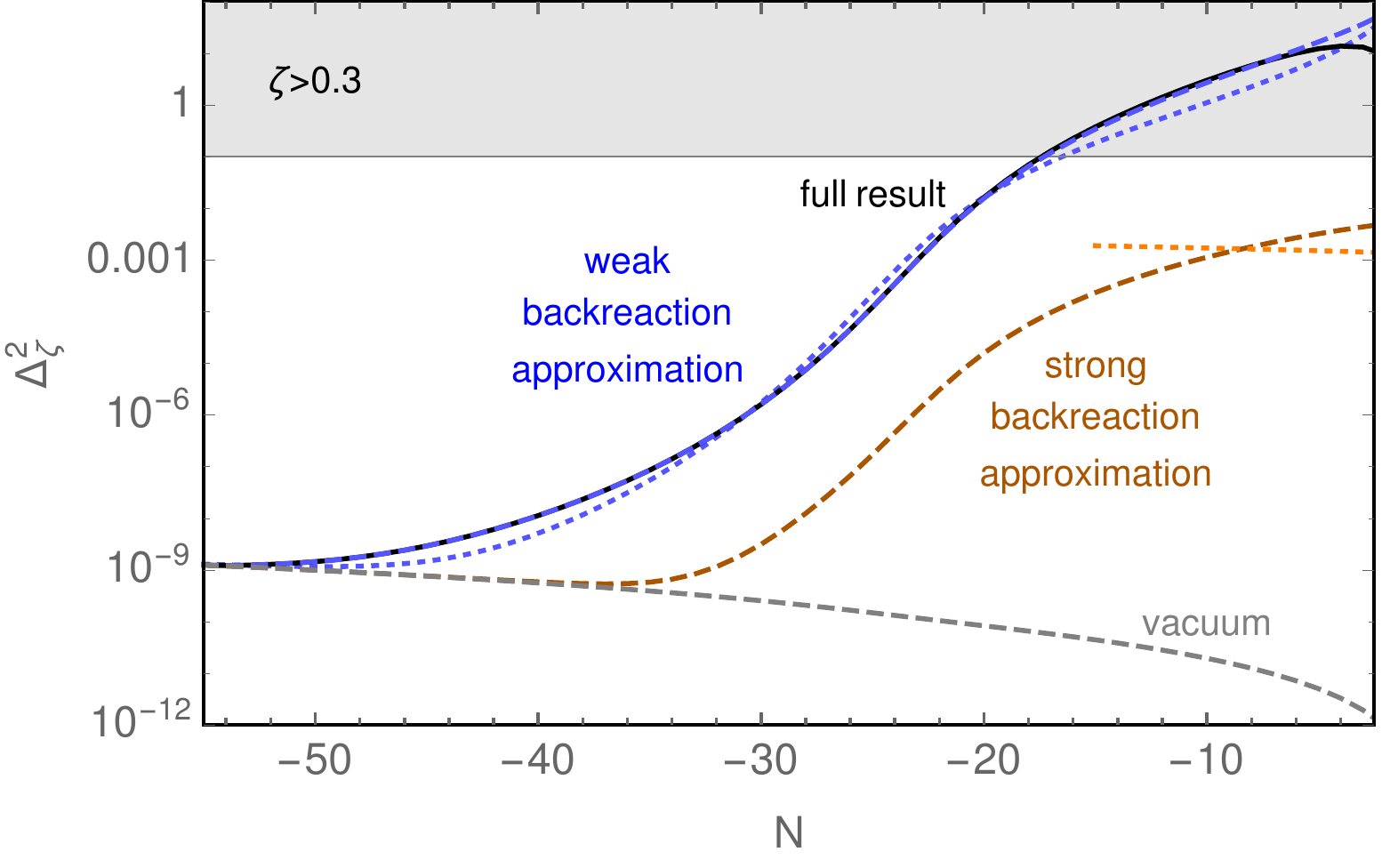}
 \caption{Scalar power spectrum sourced by Eq.~\eqref{eq:EB_estimate} for $1/f = 35$. The black curve is our full result, the dashed blue and orange curves implement the weak and strong backreaction approximation of Refs.~\cite{Barnaby:2011vw} and \cite{Anber:2009ua}, respectively. The corresponding dotted curves indicate the very good agreement with the final expressions for the power spectrum derived in these references. In this appendix we use the convention that inflation ends at $N = 0$.}
 \label{fig:powerspectrum_comparison_1}
\end{figure}

We start from the equation of motion for the scalar perturbations, Eq.~\eqref{eq:deltaphi_1},
\begin{align}
 \delta \phi'' + 3 \, \delta \phi'   - \frac{N_{,\phi}}{f H^2} \frac{\partial \langle \vec E \vec B \rangle}{\partial N} \delta \phi  =  \frac{1}{f H^2} \delta_{EB}  \,.
 \label{eq:deltaphi_1_app}
\end{align}
Ref.~\cite{Barnaby:2011vw} focuses on the regime of weak or mild backreaction (wb) where the $\partial \langle \vec E \vec B \rangle / \partial N$ term can be neglected,\footnote{We note that Eq.~\cite{Barnaby:2011vw} includes the slow-roll suppressed mass term for $\delta \phi$ and (working in Fourier space) the unequal time correlations in $\langle \delta_{EB}(N) \delta_{EB}(N') \rangle$. However, as the very good agreement in Fig.~\ref{fig:powerspectrum_comparison_1} shows, these do not significantly change the result.}
\begin{align}
 L_{N}^{(wb)}[\delta \phi(N)] \equiv \delta \phi'' + 3 \, \delta \phi'   \simeq  \frac{1}{f H^2} \delta_{EB}  \,.
 \label{eq:deltaphi_1_wb} 
\end{align}
Following the steps in Eq.~\eqref{eq:deltaphi_1} to \eqref{eq:powerspectrum_et} of the main text yields
\begin{align}
 \langle \delta N^2 \rangle^{(wb)} \simeq N_{,\phi}^2 \int dN' \frac{G_{wb}^2(N,N') \sigma_{EB}^2(N')}{f^2 H^2(N')}\,,
 \label{eq:powerspectrum_wb}
\end{align}
with $G_{wb}(N,N')$ denoting the Greens function of the linear operator $L_{N}^{(wb)}$. 

Ref.~\cite{Anber:2009ua} focuses on the opposite limit of strong backreaction. In this case, the  the backreaction term in Eq.~\eqref{eq:deltaphi_1_app} can be approximated as
\begin{align}
 \frac{N_{,\phi}}{f H^2} \frac{\partial \langle \vec E \vec B \rangle}{\partial N} \delta \phi \simeq \frac{1}{2 f^2 H^2} \frac{\partial \langle \vec E \vec B \rangle}{\partial \xi} \delta \phi' \simeq 
 \frac{1}{2 f^2 H^2} \left(2 \pi \langle \vec E \vec B \rangle \right) \delta \phi' \simeq 
 \frac{2 \pi}{2 f H^2} V_{,\phi} \delta \phi' \,.
 \label{eq:approximations}
\end{align}
In the first step, we have Taylor expanded $\langle \vec E \vec B \rangle$ in terms of $\xi$ instead of $N$. This is valid if $\langle \vec E \vec B \rangle$ can be expressed as a function of $\xi$ only and if $\xi$ is strictly monotonic, implying that the evolution of $\xi$ can serve as a well-defined `clock' during inflation. As long as the fluctuations are small, $\delta N, \delta \xi \ll 1$, both descriptions are then equivalent. In the full system studied in the main text where $\xi$ becomes an oscillating function, this procedure can not be applied. The second step relies on the explicit form of $\langle \vec E \vec B \rangle$ in Eq.~\eqref{eq:EB_estimate} with the additional assumption of $H$ being approximately constant. The final step uses the background equation of motion in the strong backreaction regime where the $\dot \phi$-term can be neglected.\footnote{In our numerical evolution of this system of $1/f = 35$ we find all three terms of the background eom to be of similar size towards the end of inflation. This approximation thus induces an ${\cal O}(5)$ error in the Greens function, which is squared in the power spectrum and essentially accounts for the discrepancy between the black and dashed orange curve.} Based on this, Eq.~\eqref{eq:deltaphi_1_app} can be expressed as
\begin{align}
 L_N^{(sb)} [\delta \phi(N)] \equiv \delta \phi'' + 3 \delta \phi' - \frac{\pi}{f H^2} V_{,\phi} \delta \phi' \simeq \frac{1}{f H^2} \delta_{EB} \,,
 \label{eq:deltaphi_1_sb}
\end{align}
and correspondingly
\begin{align}
 \langle \delta N^2 \rangle^{(sb)} \simeq N_{,\phi}^2 \int dN' \frac{G_{sb}^2(N,N') \sigma_{EB}^2(N')}{f^2 H^2(N')}\,,
 \label{eq:powerspectrum_sb}
\end{align}
with $G_{sb}(N,N')$ denoting the Greens function of the linear operator $L_{N}^{(sb)}$.

Fig.~\ref{fig:powerspectrum_comparison_1} compares our formalism (black curve) with the approximations performed in Ref.~\cite{Barnaby:2011vw} (blue curves) and Ref.~\cite{Anber:2009ua} (orange curve). In all cases, for the purpose of the comparison with previous results, we assume in this appendix $\langle \vec E \vec B \rangle$ to be given by Eq.~\eqref{eq:EB_estimate} and correspondingly $\sigma^2_{EB} \simeq \langle \vec E \vec B \rangle$ (see e.g.\ Ref.~\cite{Linde:2012bt}). The black solid curve indicates our result based on \eqref{eq:powerspectrum_et}, i.e.\ including the gauge field backreaction in the $\delta \phi$ equation of motion, with the gray dashed curve displaying for reference the vacuum contribution. The dashed blue curve (essentially coinciding with the black curve) is the result obtain based on the linear operator~\eqref{eq:deltaphi_1_wb} in the weak backreaction regime, the dashed orange curve is correspondingly based on the linear operator~\eqref{eq:deltaphi_1_sb} in the strong backreaction regime\footnote{{Note that the strong backreaction approximation can only be expected to be valid at large values of $\xi$, towards the end of inflation.}}. The dotted blue and orange curves are the results derived in Refs.~\cite{Barnaby:2011vw} and \cite{Anber:2009ua} for the weak and strong backreaction regime, respectively, demonstrating our ability to reproduce these results when using the same approximations. Finally, in the gray shaded region $\zeta \geq 0.3$, indicating that we cannot trust the perturbative analysis underlying our computations.

The excellent agreement between our full result (black) and the weak backreaction approximation (blue) indicates that the backreaction term in the $\delta \phi$ equation of motion  is essentially irrelevant for the parameters discussed here. This conclusion is in contradiction to the conclusion drawn in \cite{Anber:2009ua,Linde:2012bt}, which would indicate that backreaction dominates roughly above the dotted orange horizontal line in Fig.~\ref{fig:powerspectrum_comparison_1}, consequently suppressing the resulting power spectrum. We can track this difference down to the approximations performed in Eq.~\eqref{eq:approximations}, in particular in the last step thereof. We conclude that the sourced scalar power spectrum is two to three orders of magnitude larger than previously estimated. Nevertheless, our procedure also entails approximations which need to be scrutinized, most notably the omission of the gradients $\nabla \Phi$ and the
dropping the unequal time contribution of the $\delta_{EB}$ two-point correlator. Given the importance of this result for the production of primordial black holes, this clearly calls for further investigation.

Finally, Ref.~\cite{Linde:2012bt} presents a simplified derivation of the results obtained in Refs.~\cite{Anber:2009ua,Barnaby:2011vw}. In the strong backreaction regime this relies on the same approximations as \cite{Anber:2009ua}, hence it is not surprising that Ref.~\cite{Linde:2012bt} also finds a strong suppression of the power spectrum  in the strong backreaction regime.

\bibliographystyle{JHEP}
\bibliography{refs}{}




\end{document}